\documentclass[11pt]{article}
\pdfoutput=1
\usepackage{jcapmod} 
\usepackage{booktabs} 
\usepackage[english]{babel}												
\usepackage{amsmath,amssymb,amsbsy,amstext, amsthm, simplewick}
\usepackage{wrapfig}
\usepackage{hyperref}
\usepackage{graphicx}
\usepackage{amsfonts}
\usepackage{amssymb}
\usepackage{subfig}
\usepackage{enumitem}
\usepackage{upgreek}
\usepackage{exscale,relsize}
\usepackage[makeroom]{cancel}
\usepackage{soul}
\usepackage{bbold}
\usepackage{lipsum}
\usepackage{mdframed}
\usepackage{mathtools}
\usepackage[export]{adjustbox}
\usepackage[outdir=./]{epstopdf}
\usepackage{placeins}
\usepackage{afterpage}
\usepackage{float}
\usepackage{comment}

\allowdisplaybreaks[1]

\usepackage{colortbl}
\definecolor{lightgreen}{cmyk}{0.2, 0, 0.2, 0.2}
\definecolor{lightgray}{cmyk}{0.1,0.2,0,0.1}
\definecolor{lightgray2}{cmyk}{0.1,0.1,0,0.1}

\makeatletter
\newlength{\apb@width}
\newcommand{\autoparbox}[2][c]{\settowidth{\apb@width}{#2}\parbox[#1]{\apb@width}{#2}}

\newcommand{\Cen}[2]{%
	\ifmeasuring@
	#2%
	\else
	\makebox[\ifcase\expandafter #1\maxcolumn@widths\fi]{$\displaystyle#2$}%
	\fi
}
\makeatother

\numberwithin{equation}{section} 

\def\l{\left}
\def\r{\right}
\def\({\l(}
\def\){\r)}
\def\[{\l[}
\def\]{\r]}
\def\pd{\partial}

\def\b{\boldsymbol}

\def\til{\tilde}

\def\la{\langle}
\def\ra{\rangle}

\newcommand{\Beq}{\begin{equation}\begin{aligned}}
\newcommand{\Eeq}{\end{aligned}\end{equation}}

\newcommand{\bx}{{\bf x}}
\newcommand{\phiosc}{\phi_{\textrm{osc}}}
\newcommand{\Vnl}{V_{\textrm{nl}}}
\newcommand{\Unl}{U_{\textrm{nl}}}
\newcommand{\tsS}{\tilde{\mathcal{S}}}
\let\oldRightarrow\Rightarrow
\renewcommand{\Rightarrow}{\quad\oldRightarrow\quad}

\DeclareSymbolFont{extraup}{U}{zavm}{m}{n}
\DeclareMathSymbol{\varheart}{\mathalpha}{extraup}{86}
\DeclareMathSymbol{\vardiamond}{\mathalpha}{extraup}{87}

\begin{document}

\hypersetup{pageanchor=false}
\begin{titlepage}
	
\setcounter{page}{1} \baselineskip=15.5pt \thispagestyle{empty}
	
\bigskip\
	
\vspace{1cm}
\begin{center}
		
{\fontsize{20.74}{24}\selectfont  \sffamily \bfseries  Classical Decay Rates of Oscillons}

\end{center}
	
\vspace{0.2cm}
	
\begin{center}
{\fontsize{12}{30}\selectfont  Hong-Yi Zhang$^{\clubsuit}$\footnote{hongyi@rice.edu}, Mustafa A. Amin$^{\clubsuit}$\footnote{mustafa.a.amin@rice.edu}, \\Edmund J. Copeland$^{\spadesuit}$\footnote{ed.copeland@nottingham.ac.uk}, Paul M. Saffin$^{\spadesuit}$\footnote{paul.saffin@nottingham.ac.uk} \& Kaloian D. Lozanov$^{\vardiamond}$\footnote{klozanov@mpa-garching.mpg.de} }
\end{center}
\begin{center}
		
\vskip 7pt
		
\textsl{$^{\clubsuit}$ Department of Physics \& Astronomy, Rice University, Houston, Texas 77005, U.S.A.}\\
\textsl{$^{\spadesuit}$ School of Physics and Astronomy, University of Nottingham, Nottingham, NG7 2RD, U. K.}\\
\textsl{$^{\vardiamond}$ Max Planck Institute for Astrophysics, Karl-Schwarzschild-Str. 1, 85748 Garching, Germany}\\

\vskip 7pt
		
\end{center}
	
\vspace{1.2cm}
\hrule \vspace{0.3cm}
\noindent {\sffamily \bfseries Abstract} \\[0.1cm]
Oscillons are extremely long-lived, spatially-localized field configurations in real-valued scalar field theories that slowly lose energy via radiation of scalar waves. Before their eventual demise, oscillons can pass through (one or more) exceptionally stable field configurations where their decay rate is highly suppressed. We provide an improved calculation of the non-trivial behavior of the decay rates, and lifetimes of oscillons. In particular, our calculation correctly captures the existence (or absence) of the exceptionally long-lived states for large amplitude oscillons in a broad class of potentials, including non-polynomial potentials that flatten at large field values.  The key underlying reason for the improved (by many orders of magnitude in some cases) calculation is the systematic inclusion of a spacetime-dependent effective mass term in the equation describing the radiation emitted by oscillons (in addition to a source term). Our results for the exceptionally stable configurations, decay rates, and lifetime of  large amplitude oscillons (in some cases $\gtrsim 10^8$ oscillations) in such flattened potentials might be relevant for cosmological applications.
\vskip 10pt
\hrule
\vskip 10pt

\vspace{0.6cm}
\end{titlepage}

\hypersetup{pageanchor=true}

\tableofcontents
\section{Introduction}
Exceptionally long-lived, spatially-localized and oscillatory field configurations, called {\it{oscillons}},  exist in real-valued scalar field theories with attractive self-interactions \cite{Bogolyubsky:1976yu,Gleiser:1993pt,Copeland:1995fq,Kasuya:2002zs,Amin:2010jq}. Oscillons emerge naturally from rather generic initial conditions making them relevant for wide ranging physical contexts including reheating after inflation \cite{Amin:2010xe,Amin:2010dc,Amin:2011hj,Gleiser:2011xj,Lozanov:2017hjm,Hong:2017ooe} and other phase transitions \cite{Farhi:2007wj,Gleiser:2010qt,Bond:2015zfa}, moduli field dynamics in the early universe \cite{Antusch:2017flz}, and  structure formation in scalar field dark matter \cite{Kolb:1993hw,Olle:2019kbo,Arvanitaki:2019rax,Kawasaki:2019czd}. Oscillons can have gravitational  implications in the form of clustering \cite{Amin:2019ums}, gravitational waves \cite{Zhou:2013tsa,Antusch:2016con, Liu:2017hua, Lozanov:2019ylm,Amin:2018xfe} and even formation of primordial black holes \cite{Cotner:2019ykd,Kou:2019bbc}. They can also have non-gravitational  implications, for example in the generation of matter-antimatter asymmetry \cite{Lozanov:2014zfa}. Besides single scalar fields with canonical kinetic terms, oscillons can be found in theories with non-canonical kinetic terms \cite{Amin:2013ika,Sakstein:2018pfd} as well as multi-field systems beyond scalar fields \cite{Graham:2006vy,Gleiser:2008dt,Sfakianakis:2012bq}.\footnote{When gravity is more important than scalar-field self-interactions, oscillons are called ``oscillatons" \cite{UrenaLopez:2001tw,Alcubierre:2003sx,Ikeda:2017qev}. Oscillons are also intimately connected with Q-balls \cite{Coleman:1985ki,Nugaev:2019vru}, and boson stars \cite{Kaup:1968, Liebling:2012fv} which are related configurations in complex valued fields (without and with gravity respectively). Oscillons also have non-relativistic analogs in Bose-Einstein condensates \cite{2017Sci...356..422N}, as well as in the non-relativistic, and weak field gravity regime in astrophysical contexts \cite{Schive:2014dra, Levkov:2018kau,Niemeyer:2019aqm}.}

The longevity, and decay rates of oscillons has long been a subject of interest. A decade after the discovery of oscillons (initially called ``pulsons" \cite{Bogolyubsky:1976yu}), Kruskal and Segur provided an estimate of their exceptionally suppressed decay rates in the small amplitude limit \cite{Segur:1987mg} (also see,  \cite{Fodor:2008du,Fodor:2009kf,Fodor:2019ftc}). However, oscillons of interest in cosmology do not have small amplitudes because there exists a long-wavelength instability in small amplitude oscillons in 3+1 dimensions, whereas larger amplitude ones are safe from such long-wavelength instabilities (see, for example \cite{Amin:2010jq}).\footnote{For shorter wavelength instabilities in the small amplitude limit, which are related to quantum instabilities, see for example, \cite{Hertzberg:2010yz}.} Moreover, for many potentials relevant for  cosmology, polynomial approximations to the potential are not-sufficient (for example, in the context of inflationary physics \cite{Amin:2011hj}). The characteristics of the radiation from oscillons in non-polynomial, flattened potentials was explored numerically in \cite{Salmi:2012ta}. 

Recently, systematic, semi-analytic studies have been undertaken to understand the longevity of oscillons by taking advantage of an observation that oscillons are (partially) analogous to Q-balls \cite{Kasuya:2002zs} in that they have an approximately conserved charge in the non-relativistic limit \cite{Mukaida:2016hwd,Ibe:2019vyo,Eby:2018ufi}. While not strictly a small amplitude analysis, a small amplitude is often assumed in practice. In particular, an effective mass term in the equations of motion related to radiation is ignored \cite{Mukaida:2016hwd,Ibe:2019vyo,Eby:2018ufi}, and usually a low order polynomial is used for the scalar field potential when comparing with numerics \cite{Mukaida:2016hwd,Ibe:2019vyo}. In the context of lower order polynomial potentials, this technique elegantly captured non-monotonic behavior of the decay rate with the approximate ``charge" of the oscillon -- thus explaining the step-like behavior of adiabatic invariants associated with the oscillons as a function of time \cite{Mukaida:2016hwd,Ibe:2019vyo}.  

What was not clear to us is whether such techniques successfully capture the decay rates of large amplitude oscillons in general potentials. Evidence exists that the above mentioned techniques can fail in a particular case \cite{Ibe:2019lzv}. Motivated by, and building upon earlier works, in this paper we provide an improved calculation of the decay rate of oscillons in 3+1 dimensions. Our technique is applicable to large amplitude oscillons in polynomial and non-polynomial potentials. Crucially, by including the effects of a spacetime dependent effective mass, we are able to capture the decay rates accurately -- an improvement by many orders of magnitude in certain cases compared to earlier techniques. Our results match well with detailed numerical simulations. 

We capture the existence of deep dips in the decay rate as a function of oscillon parameters. Such dips, when present, account for most of the lifetime of the oscillons. We find that such dips are linked with unusual suppression of the radiation at the lowest kinematically allowed frequency.\footnote{In \cite{Gleiser:2009ys}, the authors proposed an analytical theory of oscillon lifetime based on the assumption that the radiation is predominantly emitted at frequencies just above the mass threshold. This turned out to be incorrect, as shown in \cite{Salmi:2012ta} --  the radiation is actually dominated by the first non-zero multiple of fundamental frequency, namely $3\omega$ in symmetric potentials and $2\omega$ in asymmetric ones.} In the cases we have explored, for certain models, oscillons can last for upwards of $10^8$ oscillations. We also find agreement of our analytic results with some numerical results for exceptionally long lifetimes quoted in the literature \cite{Olle:2019kbo}, and are also consistent with fact that 3+1 dimensional oscillons in the sine-gordon model last for less than  $10^3$ oscillations.

The rest of the paper is organized as follows. In Sec.~\ref{sec:setup}, we provide the general theoretical setup of our calculation. In Sec.~\ref{sec:profile} we provide an algorithm for calculating the single frequency oscillon profile, as well as discuss conditions for existence, uniqueness and (long-wavelength) stability of such configurations. In Sec.~\ref{sec:perturbation}, we derive and solve the equation of motion for the radiation modes including a spacetime-dependent effective mass. We provide expressions for the classical decay rates in Sec.~\ref{sec:decay_rate}. We then discuss the numerical setup for evolving the oscillon  field configurations, and their decay rates in Sec.~\ref{sec:numerics}. In Sec.~\ref{sec:examples} we compare the numerical and analytical results for a large class of models. In Sec.~\ref{sec:future} we discuss the limitations of our approach and future directions, before finally summarizing our work in Sec.~\ref{sec:conclusion}. In the appendix we collect results regarding mathematical details related to the radiation modes.

\section{Theoretical Setup}\label{sec:setup}
We begin with an action for a real-valued scalar field:\footnote{We use the $+---$ signature, and use units with $\hbar=c=1$.}
\begin{figure}[t]
	\centering
	\includegraphics[width=0.7\linewidth]{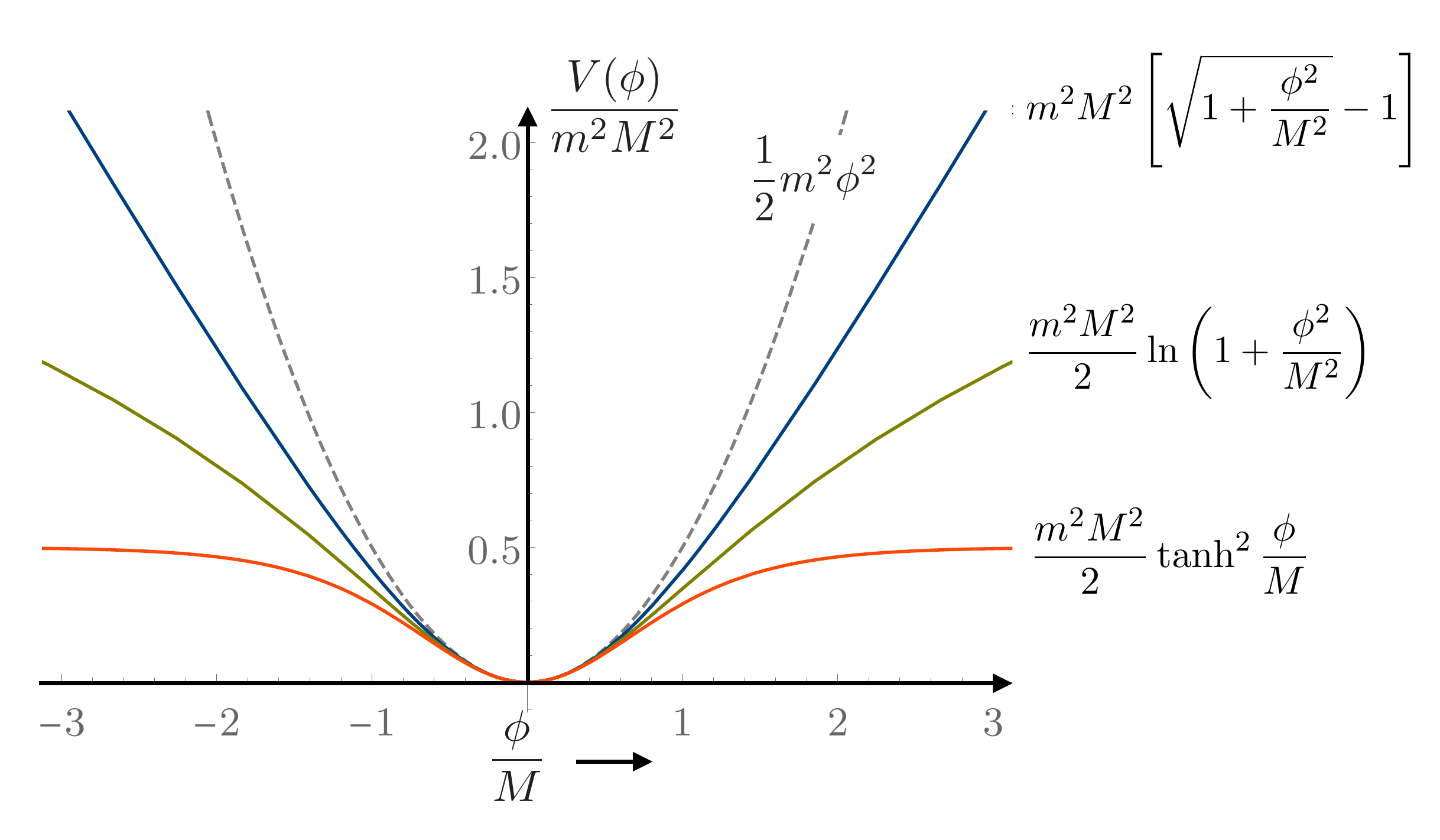}
	\caption{A sample of profiles that can support long-lived oscillons. We show potentials that open up away from the minimum, with the dashed curve showing the quadratic potential for reference. The solid lines represent potentials with quadratic minima which asymptote to different power laws of the field at large field values. The scale $M$ sets the transition from quadratic to non-quadratic behavior.}
	\label{fig:potential}
\end{figure}
\Beq
S=\int d^4 x \left(\frac{1}{2}\partial_\mu \phi \partial^\mu \phi - V(\phi)\right)\,.
\Eeq

Unless otherwise stated, we will assume that $V(\phi)=V(-\phi)$, with a single global minimum at $\phi=0$. The potential $V(\phi)$ is assumed to have a quadratic minimum, and for most of the cases considered in this paper, $V(\phi)$ will asymptote to some power law $\phi^\alpha$ with $\alpha<2$ at large field values.\footnote{This will not be true when we consider low order polynomial potentials of the form $V(\phi)=(1/2)m^2\phi^2-(\lambda /4) \phi^4+(g/6)\phi^6$ to connect with earlier literature.} The transition between the quadratic minimum and shallower than quadratic asymptotic behavior is determined by a scale $\phi\sim M$. Typically, we will be interested in the field amplitudes $\phi \lesssim \textrm{few}\times M$. Examples of such potentials are shown in Fig.~\ref{fig:potential}.

\begin{figure}[t]
	\centering
	\includegraphics[width=0.65\linewidth]{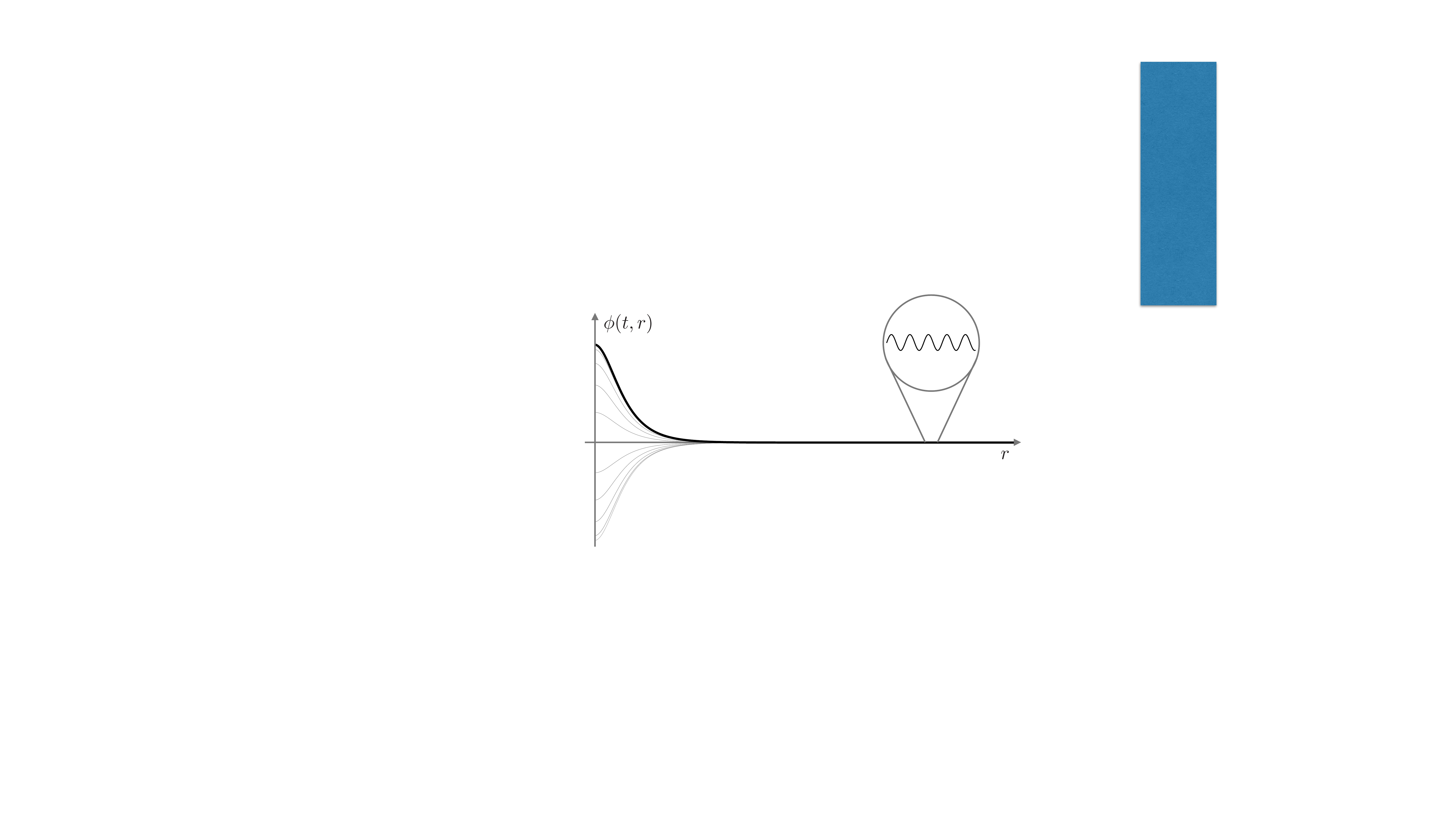}
	\caption{A schematic plot of an oscillon: a spatially-localized, oscillating field configuration and its (small) radiation tail.}
	\label{fig:oscillon}
\end{figure}

It is useful to write the potential $V(\phi)$ as 
\Beq
V(\phi)=\frac{1}{2}m^2\phi^2+\Vnl(\phi)\,.
\Eeq
where we do not make any assumptions regarding the relative size of the two terms. The equation of motion for the field is 
\Beq
\label{eq:KG}
&\left[\Box+m^2\right]\phi=-V_{\textrm{nl}}'(\phi)\,,
\Eeq
where $\square\equiv \pd_t^2-\nabla^2$. We are interested in the radiating tail of oscillons (see Fig.~\ref{fig:oscillon}). With this in mind, we split the solution of eq.~\eqref{eq:KG} into a spatially-localized, time-periodic and spherically symmetric $\phi_{\textrm{osc}}(t,r)$, and a perturbation $\xi(t,\bx)$:
\Beq
\phi(t,\bx)
&=\phi_{\textrm{osc}}(t,r)+\xi(t,\bx)\,.
\label{eq:split}
\Eeq

For our purposes, ``spatially localized" means that the solution dies faster than $1/r$ at spatial infinity. Plugging eq.~\eqref{eq:split} into eq.~\eqref{eq:KG} and expanding to leading order in $\xi$ we have
\Beq
\label{eq:EOM_linear_xi}
\boxed{
\left[\square +m^{2}\right] \phiosc + \Vnl'(\phiosc) + \left[\square+m^{2}+\Vnl''(\phiosc)\right] \xi = \mathcal{O}\left[\xi^{2}\right].}
\Eeq
We will discuss how to obtain $\phiosc(t,r)$ and $\xi(t,\bx)$ separately in the sections that follow. For a clutter-free discussion in those sections, we end this section with a collections of useful expansions and definitions. Their relevance will be more apparent in the subsequent sections.

We will assume (and justify) that a good approximation to $\phiosc$ is provided by the single frequency solution:
\Beq
\label{eq:SingleFreq}
\phiosc(t,r)\approx \Phi(r)\cos\omega t\,,
\Eeq
where $\Phi(r)$ is the spatially localized profile. We will discuss the conditions of the validity of this ansatz and the requirement for the existence of localized profiles $\Phi(r)$ in the next section. 
With the single frequency ansatz, $\Vnl$, $\Vnl'$ and $\Vnl''$ can all be expanded as a Fourier series in time as follows:
\Beq
\label{eq:CosSeriesExp}
\Vnl(\Phi \cos\omega t)=-\frac{1}{2}\mathcal{U}_0(\Phi)-\sum_{j=1}^\infty \mathcal{U}_j(\Phi) \cos(j \omega t)\,,\quad \mathcal{U}_j(\Phi)=-\frac{\omega}{\pi}\int_{-\pi/\omega}^{\pi/\omega} dt' \cos (j\omega t') \,\Vnl(\Phi \cos\omega t')\,,\\
\Vnl'(\Phi \cos\omega t)= -\frac{1}{2}S_0(\Phi) -\sum_{j=1}^\infty S_j(\Phi) \cos(j \omega t)\,,\quad S_j(\Phi)=-\frac{\omega}{\pi}\int_{-\pi/\omega}^{\pi/\omega} dt' \cos (j\omega t') \,\Vnl'(\Phi \cos\omega t')\,,\\
\Vnl''(\Phi \cos\omega t)=\frac{1}{2}\mathcal{E}_0(\Phi)+\sum_{j=1}^\infty \mathcal{E}_j(\Phi) \cos(j \omega t)\,,\quad \mathcal{E}_j(\Phi)=\frac{\omega}{\pi}\int_{-\pi/\omega}^{\pi/\omega} dt' \cos (j\omega t') \,\Vnl''(\Phi \cos\omega t')\,.
\Eeq
 Since $\Vnl(\phi)=\Vnl(-\phi)$, $\mathcal{U}_j=\mathcal{E}_j=0$ when $j$ is odd. Similarly, $S_j=0$ when $j$ is even. As we will see below, it is useful to define a $\Unl(\Phi)$:
\Beq
\label{eq:Unl}
\Unl(\Phi)\equiv-\frac{1}{2}\mathcal{U}_0(\Phi)=-\la \Vnl(\Phi \cos\omega t')\ra_T\,,\quad\textrm{and}\quad \Unl'(\Phi)=S_1(\Phi)/2=-\la \cos(\omega t')\Vnl'(\Phi \cos\omega t')\ra_T\,.
\Eeq
The above expansions rely on periodicity in time, but do not rely on having a convergent Taylor series expansion of the potential $V_{\rm nl}(\phi)$. In contrast to earlier work, these expansions will allow us to access amplitudes beyond the radius of convergence of the Taylor series (around $\phi=0$) -- these expansions generalize the technique described in \cite{Piette:1997hf,Mukaida:2016hwd}, and allow us to obtain the profile $\Phi(r)$ for a given $\omega$ at large amplitudes. We turn to the task of obtaining the profile $\Phi(r)$ in the next section. 

\section{Oscillon Profile} \label{sec:profile}

In this section,  we provide the equation and algorithm to solve for the profile, $\Phi(r)$, of the spatially-localized, single-frequency solution. We will also provide conditions for the existence, and uniqueness of $\Phi(r)$ for a given $\omega$, as well as the stability of $\phiosc$ (against long wavelength perturbations).

To obtain an equation for the profile $\Phi(r)$, we first substitute the single frequency solution $\phiosc=\Phi \cos\omega t$ and the cosine series expansions of $\Vnl'$ from \eqref{eq:CosSeriesExp} into eq.~\eqref{eq:EOM_linear_xi}. We now multiply eq.~\eqref{eq:EOM_linear_xi} by $\cos(\omega t)$ and integrate over a period ($2\pi/\omega$). At 0-th order in $\xi$, we obtain the {\it profile equation}:
\begin{align}\label{radial_eq}
\nabla^2\Phi - (m^2 - \omega^2)\Phi + 2\Unl'(\Phi) = 0\,,
\end{align}
where $\nabla^2=\partial_r^2+(2/r)\partial_r$ and $\Unl$ was defined in eq.~\eqref{eq:Unl}. We solve for $\Phi(r)$ as follows. For a given $\omega$, we look for a $\Phi(r=0)$ such that $\Phi(r)$ is spatially localized, regular at the origin, and has no-nodes -- a typical ``shooting" problem. Note that $\omega$ is restricted to some critical value $\omega_{\textrm{crit}}$ depending on the form of $\Unl$. 

Once such a spatially localized solution is found, its (time-averaged) energy can be calculated as follows:
\begin{align}
E_{\textrm{osc}}=&\int_0^\infty dr\, 4\pi r^2 \left\la\frac{1}{2} (\partial_t\phiosc)^2 + \frac{1}{2}(\pd_r\phiosc)^2 + \frac{1}{2}m^2\phiosc^2 + \Vnl(\phiosc)\right\ra_T \\
=&\int_{0}^{\infty}dr\, 4\pi r^2 \left[ \frac{1}{4} (\omega^2 +m^2)\Phi^{2}+ \frac{1}{4} \left(\partial_{r} \Phi\right)^{2}-\Unl(\Phi)\right]\,,
\end{align}
where $\la\cdots\ra_T=\frac{\omega}{2\pi}\int_{-\pi/\omega}^{\pi/\omega}dt(\hdots) $ denotes time-averaging over a period. 
\begin{figure}[t] 
   \centering
   \includegraphics[width=4in]{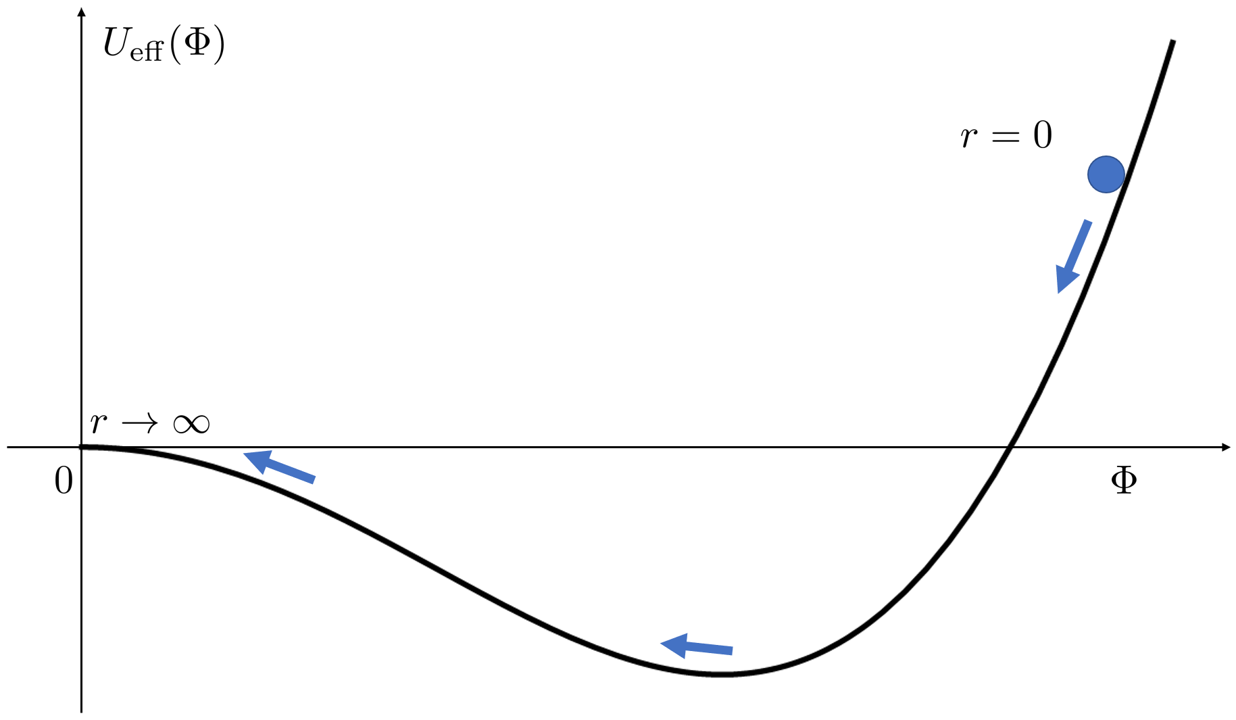} 
   \caption{\small The shape of the effective potential $U_{\textrm{eff}}(\Phi)$ for $V(\phi)$ potentials that open up away from the minimum. The profile solution can be obtained by thinking about $\Phi(r)$ as the spatial coordinate of a ball rolling down the $U_{\textrm{eff}}(\Phi)$ in the presence of friction $(2/r)\partial_r\Phi(r)$. Note that $r$ plays the role of the time coordinate.}
   \label{fig:Ueff}
\end{figure}
\subsection{Condition for Spatially Localized Profiles and Their Uniqueness}
Not all forms of $\Unl$ (and $\omega$) allow spatially localized, nodeless solutions. The necessary condition for such solutions to exist are:
\Beq
\label{profile_condition}
\frac{4\Unl(\Phi)}{\Phi^2}>m^2-\omega^2>0\,.
\Eeq
To understand why ${\Unl(\Phi)}/{\Phi^2}>(1/4)(m^2-\omega^2)$ is necessary, note that eq.~\eqref{radial_eq} can be regarded as an equation of motion for a particle with $r$ playing the role of a time variable.  Explicitly,
\Beq
\label{eq:Ueff}
\partial_r^2\Phi+\frac{2}{r}\partial_r\Phi=-U_{\textrm{eff}}'(\Phi)\,.
\Eeq
where 
\Beq
U_\textrm{eff}(\Phi)\equiv -\frac{1}{2}(m^2-\omega^2)\Phi^2 + 2\Unl(\Phi)\,.
\Eeq

We think of $(2/r) \partial_r\Phi$ as a friction term. We would like a monotonic solution with $\Phi|_{r= 0}\ne 0$, $\partial_r\Phi|_{r=0}=0$ and $\Phi,\partial_r\Phi|_{r\rightarrow\infty} \rightarrow 0$. If $\Unl(0)=0$, then the final effective energy at $r\rightarrow \infty$ is $0$.  Now, since there is friction in the system, we must have the ``initial energy" at $r=0$ satisfy $U_{\textrm{eff}}(\Phi(r=0))>0$. That is, for some $\Phi\ne 0$, we must have ${\Unl(\Phi)}>(m^2-\omega^2)\Phi^2/4$. 

To understand the second inequality, $m^2-\omega^2>0$, note that for large $r$, eq.~\eqref{eq:Ueff} has solutions of the form $\Phi|_{r\rightarrow\infty}\sim  e^{\pm i\sqrt{\omega^2-m^2}r}/r$ (assuming we can ignore $\Unl$ at large $r$ since $\Phi$ will be small). Hence, if we want localized solutions, we need $m^2>\omega^2$.
\\ \\
\noindent{\bf Uniqueness of the Profile}: For $V(\phi)$ that have quadratic minima, and  flatten to shallower than quadratic power laws at larger field values, $U_{\textrm{eff}}(\Phi)$ will be negative for small $\Phi$ and positive and monotonic for large $\Phi$ (for any $\omega<m$). Following our ball-rolling on a hill analogy with $r$ as the time coordinate, it is clear that there will be a {\it unique} initial condition $\Phi|_{r= 0}\ne 0$, $\partial_r\Phi|_{r=0}=0$, where $U_{\textrm{eff}}(\Phi|_{r=0})>0$ for which $\Phi(r)$ will be localized. 

We note that for polynomial potentials, $U_{\textrm{eff}}(\Phi)$ can have a positive local maximum at some $\Phi\ne 0$. For the case of the $\phi^6$ potential, such local maxima indicate the existence of ``flat-top" oscillons \cite{Amin:2010jq}.

Multiple studies have shown that oscillons tend to be attractors in the space of solutions (see for example \cite{Andersen:2012wg}). The existence of a unique profile for each $\omega$ allows for some freedom in setting up initial conditions for the profiles numerically. Once an approximate initial profile is set up at some sufficiently small $\omega<m$, the oscillons radiate energy quickly, and latch on to an oscillon configuration. This oscillon configuration then adiabatically passes through a unique set of subsequent oscillon configurations with slowly increasing $\omega$. The configurations continue to evolve adiabatically, emitting a small amount of radiation, until they collapse at $\omega_{\textrm{crit}}$ (discussed next). 
\subsection{Long-Wavelength Stability}
An approximately ``conserved charge", defined as
\Beq
\mathcal{N}
&=\frac{\omega}{2} \int_0^{\infty} dr 4\pi r^2 \Phi^2(r)\,,
\Eeq
 is useful in determining a critical frequency above which oscillons are unstable to long-wavelength perturbations. For oscillons to be stable against long wavelength perturbations, we need
\Beq
\frac{d\mathcal{N}(\omega)}{d\omega}<0\,, \quad{\textrm{with}}\quad \frac{d\mathcal{N}(\omega_{\textrm{crit}})}{d\omega}=0\,.\
\Eeq
This is the well known Vakhitov and Kolokolov \cite{Vakhitov:1973} stability condition. This quantity is related to the approximately conserved number of ``particles" in the oscillon or the ``charge". Also see \cite{Lee:1991ax,Nugaev:2019vru}. 
\\ \\
\noindent{\bf Caveats about the Single-Frequency Profile}: A few words regarding the single frequency assumption are in order. We assumed that a good approximation to $\phiosc$ is provided by the single frequency solution as shown in eq.~\eqref{eq:SingleFreq}.  More generally, $\phiosc(t,r)= \sum_{j=0}^{\infty}\Phi_j(r)\cos(j\omega t+\theta_j)$ can be used. For small amplitude oscillons, the profiles can be solved order by order \cite{Fodor:2009kf}. For the models (and large amplitudes) considered here, we have checked numerically that the Fourier Transform of oscillons in the temporal domain show a rich structure in other frequencies (also see \cite{Salmi:2012ta}), including frequencies other than multiples of $\omega$. Nevertheless, there is typically a single dominant frequency, and the single frequency solution remains a good approximation up to moderately large field amplitudes. We use this check to justify our single frequency approximation. The dropping of higher harmonics, however, might have consequences in the form of a somewhat larger than expected amplitude for the radiation modes $\xi$.\footnote{Heuristically, this could be because we have transferred these higher multipoles to the radiation sector in our calculation.} Also recall that even with the single-frequency assumption, we have ignored contributions from $\xi$ in the profile equation (see eq.~\eqref{radial_eq}).

\section{Radiative Perturbations}\label{sec:perturbation}

In this section, we will explicitly derive the perturbation equations and decay rate, and introduce a trick to include the impact of a spacetime-dependent source and a spacetime dependent effective mass term together.
\subsection{Equations of Motion}
To obtain the equation of motion for $\xi$, we begin by substituting the solution of eq.~\eqref{radial_eq} into eq.~\eqref{eq:EOM_linear_xi}, to obtain
\begin{align}\label{EOM_perturbation}
\left[ \pd_t^2 - \nabla^2 +m^2 + \mathcal{E}(t,r) \right] \xi(t,r) = S(t,r) ~,
\end{align}
where we have assumed spherical symmetry for the solutions, and defined the ``effective mass term" and the ``source term":
\Beq
\label{eq:EffMS}
\mathcal{E}(t,r)&\equiv\Vnl''(\Phi\cos\omega t))=\frac{1}{2}{\mathcal{E}}_0(r)+\sum_{j=2}^{\infty}\mathcal{E}_j(\Phi)\cos(j\omega t)|_{j={\rm even}}\,,\\
S(t,r)&\equiv -\Vnl'(\Phi \cos\omega t)+S_1(\Phi)\cos\omega t=-\sum_{j=3}^\infty S_j(\Phi) \cos(j \omega t)|_{j={\rm odd}}\,.
\Eeq
Recall that $\Phi=\Phi(r)$ is the single frequency profile obtained in the previous section, and the inverse Fourier transforms $\mathcal{E}_j(\Phi)$ and $S_j(\Phi)$ were defined in eq.~\eqref{eq:CosSeriesExp}. The fact that the effective mass contains only even multiples, and the source contains only odd multiples of $\omega$ is a consequence of our assumption that $\Vnl(\phi)=\Vnl(-\phi)$.

Note that we do not include $S_1(\Phi)\cos(\omega t)$ term in $S(t,r)$ since that term was included in the profile equation. Similarly, we implicitly ignore any term on the l.h.s. of eq.~\eqref{EOM_perturbation} that is proportional to $\cos\omega t$ since that is also in principle included in the profile equation (although we ignored these $\xi$ corrections in the calculation of the profile).

We are interested in solutions $\xi(t,r)$ of eq.~\eqref{EOM_perturbation} that represent outgoing radiation at spatial infinity. In absence of the effective mass term, $\mathcal{E}(t,r)=\Vnl''(\phiosc)$, it is possible to write down a general solution using a retarded Green's function of the free Klein-Gordon operator ($\Box+m^2$) convolved with the source. This is essentially what is done in \cite{Mukaida:2016hwd,Ibe:2019vyo}. However, if the space-time dependent effective mass is non-negligible compared to $m^2$, there is no simple Green's function solution. For the flattened potentials we consider, and moderately large amplitudes ($\Phi(r=0) \sim M$), the spacetime dependent effective mass $\Vnl''(M \cos\omega t)$ is of the same order as the bare mass term: $m^2$. Hence, it is not obvious that we can ignore the effective mass term. As we will see, if we want $\xi$ solutions to agree well with the numerics even qualitatively, inclusion of the effective mass term is essential.
\subsection{Separable Series Expansion}
To make progress, given the even and odd structure of the expansions of $\mathcal{E}(t,r)$ and $S(t,r)$ in eq.~\eqref{EOM_perturbation}, it is tempting to expand the solution in terms of a (separable) series in multiples of $\omega$:
\begin{align}\label{perturbation_expansion}
\xi(t,r)=\sum_{j=3}^{\infty} \xi_j(r) \cos(j\omega t)|_{j=\textrm{odd}} ~.
\end{align}
Because of the separable nature, we will be able to solve for $\xi_j(r)$. Nevertheless, the above solutions are obviously standing waves, and not the outgoing radiation modes we are after. We will extract the outgoing wave solutions from this standing wave solution as follows.  At large radii, we expect the standing wave solutions $\xi_j(r)\cos(j\omega t)$ for each $j$ will naturally split into incoming wave part $\sim r^{-1}\cos(\kappa_j r+j\omega t)$ and an outgoing wave part $\sim r^{-1}\cos(\kappa_j r-j\omega t)$ with wavenumbers 
\Beq
\kappa_j\equiv \sqrt{(j\omega)^2-m^2}\,.
\Eeq
The outgoing wave solution at large radii is obtained by simply ignoring the incoming wave part and then doubling the remaining result. Subtleties and details related to this procedure are discussed further in the Appendix \ref{sec:derivation_xi}. \footnote{This standing wave expansion is related to the small amplitude, quasi-breather expansion considered in \cite{Fodor:2009kf}, where the amplitude of outgoing waves is related to a minimization procedure. We do not carry out this minimization procedure here.}

Note that we do not include $j=1$ term in the solution for $\xi$, since at large radii, a free-traveling wave solution with frequency $j\omega =\omega<m$ does not exist (the wavenumber $\kappa =\sqrt{\omega^2-m^2}$ is imaginary). It corresponds to a localized mode.
\subsection{Perturbation Solutions}
We now proceed to solve for $\xi_j(r)$ by substituting eq.~\eqref{perturbation_expansion} in eq.~\eqref{EOM_perturbation} and collecting co-efficients of $\cos(j\omega t)$. This procedure yields
\Beq
\label{eq:xi_master}
\left[\nabla^2+\kappa_j^2\right]\xi_j=-S_j+\frac{1}{2}\sum_{l=3}^{\infty}\left(\mathcal{E}_{j+l}+\mathcal{E}_{|j-l|}\right)\xi_l\equiv -\mathcal{S}_j\,.
\Eeq
If we ignore the effective mass terms, $\mathcal{S}_j=S_j$, then we can derive a solution using the Green's function of the Helmholtz operator ($\nabla^2+\kappa_j^2$). Even without ignoring the effective mass terms, a formal (implicit) solution to this equation using the same Green's function can be obtained (see Appendix \ref{sec:derivation_xi}, or as can be checked by direct substitution)
\begin{align}
\label{eq:xi_j_general}
\xi_j(r)
&= \frac{\cos(\kappa_j r)}{\kappa_j r} \int_0^r dr' ~ \mathcal{S}_j(r')~ r' \sin(\kappa_j r') + \frac{\sin(\kappa_j r)}{\kappa_j r} \int_r^\infty dr' ~ \mathcal{S}_j(r')~ r' \cos(\kappa_j r') ~,
\end{align}
where we have ignored the homogeneous solutions of $(\nabla^2+\kappa^2)$. It is worth noting that since both $S_j$ and $\mathcal{E}_j$ are constructed out of profiles $\Phi(r)$ then, since $\Phi(r)$ is exponentially localized, we expect that the integrals in the expressions for $\xi_j$ converge rapidly at large $r$.

At $r=0$, we have
\Beq
\label{eq:xi(0)}
\xi_j(0)=\int_0^\infty dr' ~ \mathcal{S}_j(r')~ r' \cos(\kappa_j r')~.
\Eeq
This equation proves extremely useful in solving for the $\xi_j(r)$ numerically. 

First, if we assume $|\xi_j|\gg |\xi_{j+1}|$ and set $\xi_j=0$ beyond some fixed $j=N$, we have a finite set of coupled ordinary differential equations \eqref{eq:xi_master}.\footnote{We will see this in our numerical results (apart from certain special circumstances), and it was also observed by the authors in \cite{Salmi:2012ta}.} Naively solving eq.~\eqref{eq:xi_j_general} with two initial conditions at the origin will include a combination of the homogeneous and inhomogeneous solutions. To avoid the homogeneous solution, we use the following procedure.

To solve for $\xi_j(r)$ we start by constructing a trial solution by solving the $N$ coupled radial ODE equations  with the initial conditions given by $\partial_r\xi_j|_{r=0}=0$ and $\xi_j|_{r=0}=\xi_{j,0}$ for each $j\le N$. We then plug this solution into eq.~\eqref{eq:xi(0)} to see if the integral on the r.h.s matches our initial input of $\xi_j|_{r=0}=\xi_{j,0}$ on the l.h.s. If it does not, we start the process again using the result of the r.h.s of eq.~\eqref{eq:xi(0)} as our new $\xi_j|_{r=0}$.  In practice, we find that this process converges quite rapidly, and we find the unique numerical solution to eq.~\eqref{eq:xi_master}.
\subsection{Perturbations in the Large $r$ Limit}
Once the solutions $\xi_j(r)$ have been found, we can plug them into the formal expression for the solution \eqref{eq:xi_j_general}. In the limit that $r$ is large compared to the width of the profile, eq.~\eqref{eq:xi_j_general} yields
\Beq
\label{eq:xi_large_r}
\xi_j(r)
&\approx  \frac{\cos(\kappa_j r)}{\kappa_j r} \int_0^\infty dr' ~ \mathcal{S}_j(r')~ r' \sin(\kappa_j r')~,\\
\Eeq
where in the first line we replaced the limits of integration by $r\rightarrow \infty$ in eq.~\eqref{eq:xi_j_general} but kept the $r$ dependence in the coefficients. Note that we use the numerically solved $\xi_j(r')$ in $\mathcal{S}_j(r')$ on the r.h.s.

The integral on the r.h.s. of eq.~\eqref{eq:xi_large_r} is simply the Fourier Transform of $\mathcal{S}_j(r)$ evaluated at $k=\kappa_j$. That is,
\Beq
\label{eq:xi_large_r_F}
\xi_j(r)
&\approx\frac{1}{4\pi}\frac{1}{r}\cos(\kappa_j r)\tsS_j(\kappa_j)\,~,\\
\Eeq
where
\Beq
\tsS_j(k)=\int d^3 {\bf r}' \mathcal{S}_j(r')e^{-i {\bf k}\cdot {\bf r}}=\int_0^{\infty} dr'4\pi r'^2 \frac{\sin(k r')}{k r'}\mathcal{S}_j(r')~.
\Eeq
This form of $\xi_j(r)$ can be substituted into eq.~\eqref{perturbation_expansion} to get the following general solution at large $r$:
\Beq
\xi(t,r)\approx \frac{1}{8\pi r}\sum_{j=3}^{N}\tsS_j(\kappa_j)\cos(\kappa_j r-j\omega t)+\frac{1}{8\pi r}\sum_{j=3}^{N}\tsS_j(\kappa_j)\cos(\kappa_j r+j\omega t)~\,.
\Eeq
We now ignore the incoming wave part, and {\it double} the result of the outgoing waves to get 
\Beq
\label{eq:xi_rad}
\xi^{\rm rad}(t,r)\approx \frac{1}{4\pi r}\sum_{j=3}^{N}\tsS_j(\kappa_j)\cos(\kappa_j r-j\omega t)\,,
\Eeq
where $j$ is odd and recall that $\kappa_j=\sqrt{(j\omega)^2-m^2}$. Note that this solution is consistent (in the limit of vanishing effective mass) with the one obtained using the retarded Green's function of the  Klein-Gordon operator ($\Box+m^2$), only after we double the result. Also see the Appendix.

\section{Classical Decay Rate}\label{sec:decay_rate}
Given our solution \eqref{eq:xi_rad} for outgoing radiation at large radii (large compared to the width of the profile $\Phi(r)$), we can calculate the time-averaged outgoing flux as follows:
\begin{align}
&\langle T_{0r}\rangle_T=\langle\partial_t\xi^{\rm rad}(t,r)\partial_r\xi^{\rm rad}(t,r)\rangle_T = -\frac{1}{32\pi^2r^2}\sum_{j=3}^N [\tsS_j(\kappa_j)]^2{\omega_j\kappa_j}~.
\end{align}
The decay rate of oscillons is then given by
\begin{align}\label{decayrate}
\Gamma_{(N)}\equiv \frac{1}{E_{\rm osc}}\langle 4\pi r^2 T_{0r}\rangle_T= -\frac{1}{8\pi E_{\rm osc}}\sum_{j=3}^N [\tsS_j(\kappa_j)]^2\omega_j\kappa_j=\sum_{j=3}^N \Gamma_j~.
\end{align}

Recall that $\tsS_j(\kappa_j)$ is a  Fourier transform of a function $\mathcal{S}_j(r)$ constructed out of the spatially localized profile $\Phi(r)$  with a width $ w\sim {\rm few}\, m^{-1}$ (both via the source and effective mass terms). The width of the Fourier transform $\tilde{S}(\kappa)$ will be $\lesssim w^{-1}$, with an exponentially supressed amplitude for $\kappa \gtrsim w^{-1}$. As $\kappa_j\gtrsim m \gtrsim w^{-1}$, we expect $\tsS_j(\kappa_j)$ to have an exponential suppression. It is this suppression that is responsible for the small decay rate of oscillons. 

Furthermore, note that we generally expect  $[\tsS_j(\kappa_j)]^2\gg [\tsS_{j+1}(\kappa_{j+1})]^2$ because $\tsS_{j+1}$ is evaluated at a higher momentum $\kappa_{j+1}$. However, this hierarchy can be violated if some $\tsS_j(\kappa_j)$ were to vanish accidentally for some particular situation. That is, there might be an oscillon profile (specified by $\omega$) for which $\tsS_j(\kappa_j)=0$ for some particular $j$. We will see that such situations do arise when considering moderate amplitude oscillons. 

We note that it is highly unlikely that $\tsS_j(\kappa_j)$ vanish for all $j$, hence the total decay rate is always expected to be finite. Our present analysis does not lend support to conjectures of infinite lifetimes \cite{Honda:2001xg,Gleiser:2019rvw,Gleiser:2020zaj} in {\it asymmetric} double-well potentials.


\section{Numerical Setup}\label{sec:numerics}
We apply the following numerical strategy to verify our analytical results. The main goal is to solve the nonlinear Klein-Gordon equation \eqref{eq:KG} and obtain a decay rate as a function of time. Since at each instant in time we are passing adiabatically through different oscillon configurations (specified by an $\omega (t)$), these results can be directly compared to the analytically obtained decay rates from the previous sections. 

We solve the nonlinear Klein-Gordon equation \eqref{eq:KG} (assuming spherical symmetry) using a Verlet method (a 2nd-order symplectic method) while the spatial derivative is characterized by centered difference. The simulations are performed on a box of size $r_{\rm max}=60m^{-1}$ with $dt=dr/5=0.005 m^{-1}$. We have checked that changing the box size or the spatial/temporal step size does not change our results qualitatively. The size of the box is much larger than the typical width of the oscillon profile which is $<\mathcal{O}[10]\,m^{-1}$. 

At the boundary $r\rightarrow r_{\rm max}$ we impose the absorbing boundary condition \cite{Salmi:2012ta}, i.e. 
\begin{align}
\partial_{t}^{2} \phi+\partial_{t} \partial_{r} \phi+\frac{1}{r} \partial_{t} \phi+\frac{1}{2} m^{2} \phi=0 ~,
\end{align}
which uses a backward-in-time, and centered-in-space discretization. This boundary condition provides an alternative approach to remove the dispersive waves from the lattice, requiring no extra lattice sites for its operation.

We begin with spatial profile $\phi(t,r)|_{t=0}$ and $\partial_t\phi(t,r)|_{t=0}$ which is smooth at the origin $r=0$. After picking an $\omega$, we find the profile using the shooting algorithm discussed in Sec.~\ref{sec:profile}. Once set up in this way, the system evolves primarily via radiation of scalar modes which are approximately removed at $r=r_{\rm max}$. Typically the characteristic $\omega$ of the solution (near $r=0$) increases with time, and the oscillon undergoes an adiabatic evolution, passing through many oscillon configurations with increasing $\omega$. The frequency $\omega$ is measured using the interval between the field maxima at $r=0$.

The decay rate of the oscillons is numerically calculated using
\begin{align}
\Gamma(t) = \frac{1}{T_{\rm ave}} \int_{t-T_{\rm ave}/2}^{t+T_{\rm ave}/2} \frac{1}{E_{\rm osc}(t')}\frac{dE_{\rm osc}(t')}{dt'} dt' ~,
\end{align}
where we use $T_{\rm ave}=200m^{-1}$ for convenience. The time-dependent, but slowly decreasing energy $E_{\rm osc}(t)$ of the oscillon is calculated using
\Beq
E_{\rm osc}(t)=\int_0^{r_{\max}/2} dr\, 4\pi r^2 \left[\frac{1}{2} (\partial_t\phi)^2 + \frac{1}{2}(\pd_r\phi)^2 + \frac{1}{2}m^2\phi^2 + \Vnl(\phiosc)\right]\,.\\
\Eeq
Our choice of bounding radius $r_{\rm max}/2$ is arbitrary. However, as long as the bounding radius is $
\gtrsim\mathcal{O}[10]m^{-1}$, the decay rate is approximately independent of this choice.
 
We also keep track of the time averaged frequency, central amplitude, energy and decay rates of oscillons obtained by an average over a time period $T_{\rm ave}=200m^{-1}$ unless otherwise stated.

To get a more refined picture of the frequency content of the oscillons and the radiation, we calculate Fourier Transform of the time dependence of the field at $r=0$ and $r=r_{\rm rad}=50 m^{-1}$ respectively. Such Fourier Transforms are calculated over a time interval of $T_{\rm fourier}=5000m^{-1}$. 

We have confirmed that a slight change of parameters ($r_{\rm max},dt,T_{\rm ave},T_{\rm fourier}$) will not affect the results significantly.

\section{A Comparison Between Analytical and Numerical Results}\label{sec:examples}
In this section we compare the analytic and numerical results for the decay rates of oscillons for several different models (potentials).
\\ \\
\noindent{\bf Models}:  We consider potentials of the form \cite{Amin:2011hj} (motivated by Axion-Monodromy models \cite{Silverstein:2008sg,McAllister:2014mpa}):
\Beq\label{eq:PotGen}
V(\phi)=
\frac{m^2M^2}{q}\left[\left(1+\frac{\phi^2}{M^2}\right)^{q/2}-1\right]\,.
\Eeq
The smooth transition from a quadratic potential, to some shallower than quadratic region will be parametrized by a scale $M$, so that for $\phi\ll M$ we have a quadratic potential, whereas for $\phi\gg M$, the potential asymptotes to a shallower than quadratic form:

\begin{equation}
V(\phi)
=
\begin{cases}
\dfrac{1}{2}m^2\phi^2&|\phi|\ll M\,,\\ \\
\dfrac{m^2M^2}{q}\left(\dfrac{\phi^2}{M^2}\right)^{q/2}&|\phi|\gg M \quad \textrm{and}\quad0<q<2,\\ \\
-\dfrac{m^2M^2}{q}&|\phi|\gg M \quad \textrm{and}\quad q <0\,.\\
\end{cases}
\end{equation}
Other examples of potentials we consider, include
\Beq
V(\phi)=\frac{m^2M^2}{2}\tanh^2\frac{\phi}{M}\qquad\textrm{and}\qquad V(\phi)=m^2M^2\left[1-\cos\frac{\phi}{M}\right]\,.
\Eeq
The $\tanh^2$ potential is part of the $\alpha$-attractor models of inflation \cite{Kallosh:2013hoa,Lozanov:2017hjm}, whereas the cosine potential is characteristic of ``usual" axions. For the cosine potential we limit ourselves to $|\phi|<\pi M$. Some examples of the types of potentials we consider are shown in Fig.~\ref{fig:potential}. 
\\ \\
\noindent{\bf Analytics}: Within each model, we obtain a decay rate as a function of $\omega$ from the analytic calculation (where $\omega$ parametrizes a continuous set of oscillon profiles).\footnote{Note that when referring to analytic results, we allow for numerical solutions of the profile $\Phi(r)$ and the $\xi_j(r)$, which are obtained via shooting.} 
 This analytically calculated decay rate can be further broken down into the contributions from radiation at different multiples of $\omega$, with the total decay rate defined in \eqref{decayrate}, $\Gamma_{(N)}=\sum_{j=3}^N \Gamma_j$. The upper limit $N$ denotes the highest multiples of $\omega$ to be included. For example $\Gamma_{(3)}$ only includes radiation with frequency $3\omega$, whereas $\Gamma_{(5)}$ includes radiation at $3\omega$ and $5\omega$.  Also recall that for any $N$, $\Gamma_{(N)}$ includes the contribution for the effective mass since, from \eqref{eq:xi_master} we have $\mathcal{S}_j=S_j-\frac{1}{2}\sum_{l=3}^{N}\left(\mathcal{E}_{j+l}+\mathcal{E}_{|j-l|}\right)\xi_l$ (in position space, where $\mathcal{E}_j$ arise from the effective mass term). For comparison, if we ignore the effective mass terms, then we will use the notation $\Gamma^{\rm{old}}_{(N)}$. 
 
For ease of reference, we write down the decay rates for $N=3$ and $5$ explicitly below:
 \Beq
\Gamma_{(3)}&=\Gamma_3=-\frac{1}{8\pi E_{\rm osc}}[\tsS(\kappa_3)]^2(3\omega)\kappa_3\,,\\
\Gamma_{(5)}&=\Gamma_3+\Gamma_5=-\frac{1}{8\pi E_{\rm osc}}[\tsS(\kappa_3)]^2(3\omega)\kappa_3-\frac{1}{8\pi E_{\rm osc}}[\tsS(\kappa_5)]^2(5\omega)\kappa_5\,.
\Eeq
where $\tsS$ is the spatial Fourier Transform of $\mathcal{S}_j$.
Note that if $\tsS(\kappa_3)$ vanishes for some $\omega$, then $\Gamma_{(3)}$ also vanishes. However, for the same $\omega$, we will typically have $\Gamma_{(5)}=\Gamma_5\ne 0$. 
\\ \\
\noindent{\bf Numerics}: We numerically evolve the nonlinear Klein-Gordon equation \eqref{eq:KG} (assuming spherical symmetry) and calculate the decay rate as a function of time. This time dependence of the decay rate is translated to an $\omega$ dependence since the solution evolves slowly, and continuously through different oscillon configurations (characterized by an adiabatically changing $\omega(t)$).

We typically start the calculation by evolving the nonlinear Klein-Gordon equation \eqref{eq:KG} with field configurations corresponding to $\omega$ that are smaller than the ones shown in the upcoming plots. Regardless of the starting points, we always end up on the same $\Gamma-\omega$ trajectory numerically. This is a consequence of oscillons being attractors in the space of solutions, and the fact that there is a unique oscillon profile for each $\omega$.
\begin{figure}[t]
	\centering
	\begin{minipage}{0.45\linewidth}
		\includegraphics[width=\linewidth]{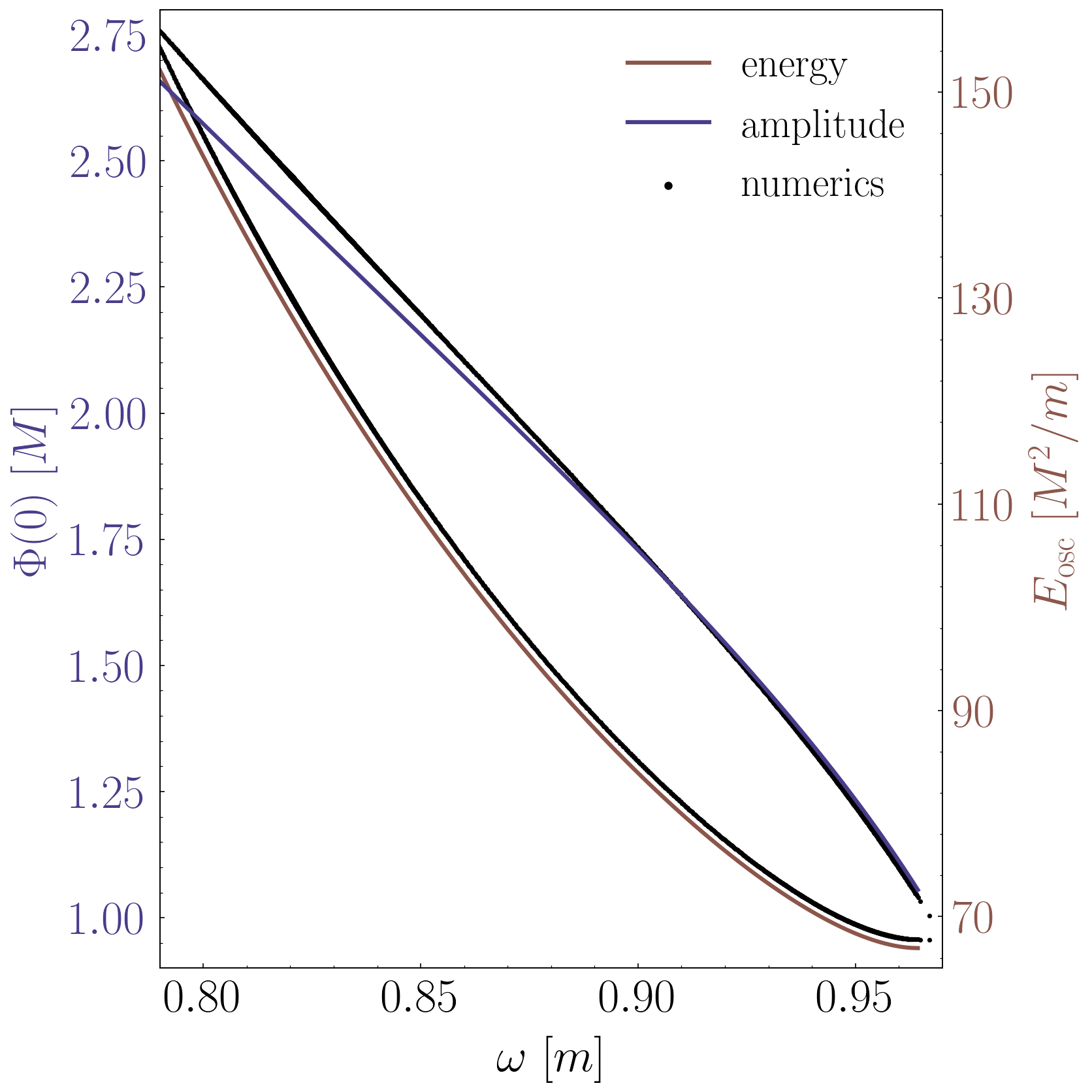}
	\end{minipage}
	\qquad
	\begin{minipage}{0.45\linewidth}
		\includegraphics[width=\linewidth]{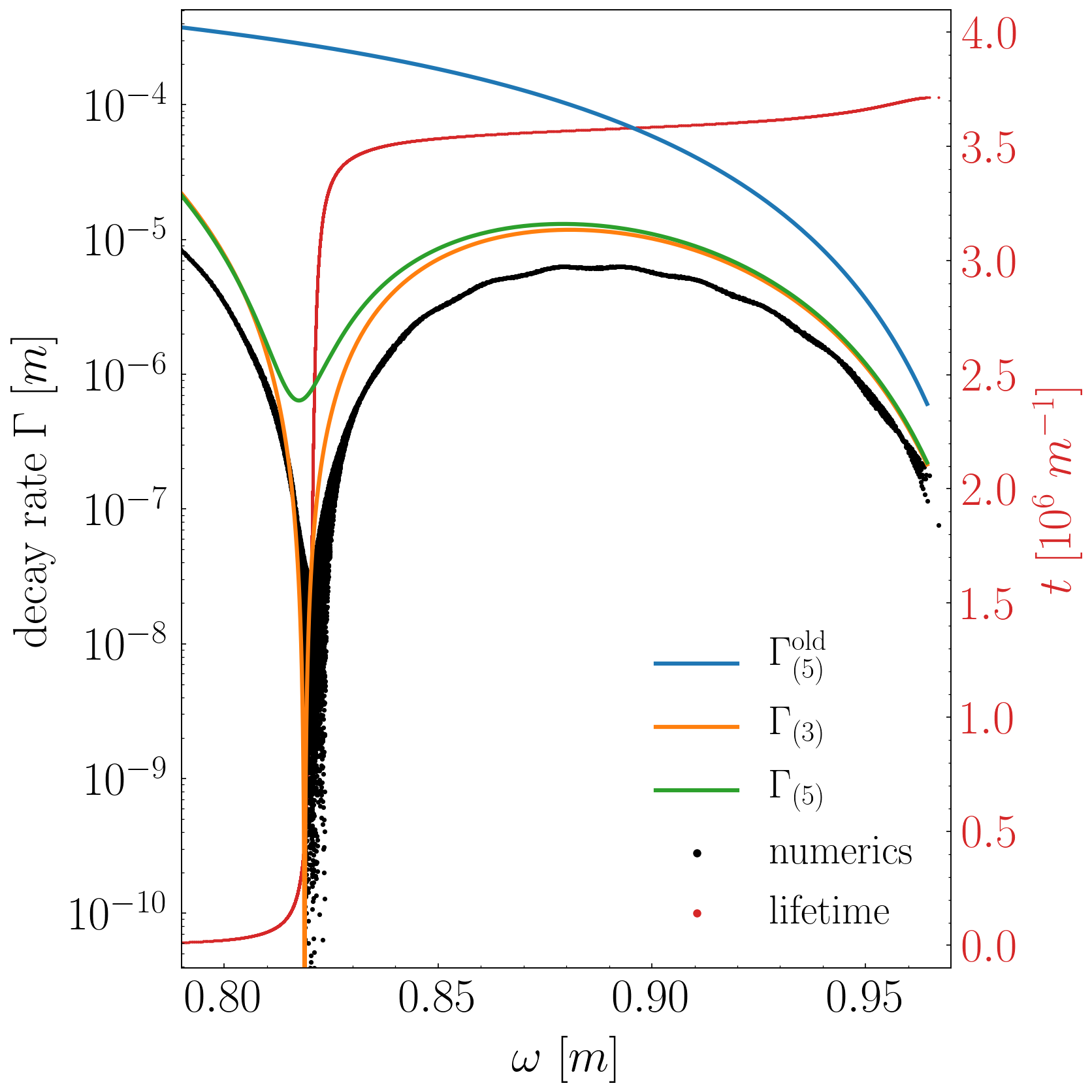}
	\end{minipage}
	\caption{\small{$V(\phi)=(1/2)m^2M^2\tanh^2(\phi/M)$: {\it Left}: Analytical (colored) and numerical (black) calculations for the oscillon amplitude and energy as a function of the fundamental oscillon frequency $\omega$. The numerical calculation includes time evolution (moving from left to right), whereas the analytical one assumes a stationary configuration for each $\omega$. {\it Right}: Decay Rate of oscillons as a function of $\omega$. Black dots show the numerical evolution of the decay rate (with time flowing from left to right, the oscillons disappear quickly after $\omega_{\rm crit}\approx 0.964m$, where $\omega_{\rm crit}$ is defined in \eqref{profile_condition}). The orange and green curves show the analytic expectation for the decay rate at each quasi-stable oscillon configuration. The orange curve includes the $3\omega$ radiation contribution, whereas the green includes the $3\omega$ and $5\omega$ modes. For comparison, the blue curve ignores the contribution from the effective mass -- and provides a much poorer estimate. A significantly improved magnitude of the decay rate, including the dip where the $3\omega$ radiation is vanishing, is correctly provided by our calculations. Finally, the red line shows that the oscillon spends most of its lifetime near the dip at $\omega_\star\approx 0.82m$.}}
	\label{fig:tanhfreqenergyamp}
\end{figure}

\subsection{The Hyperbolic Tangent Potential}
\begin{figure}[t]
	\centering
	\begin{minipage}{0.496\linewidth}
		\includegraphics[width=\linewidth]{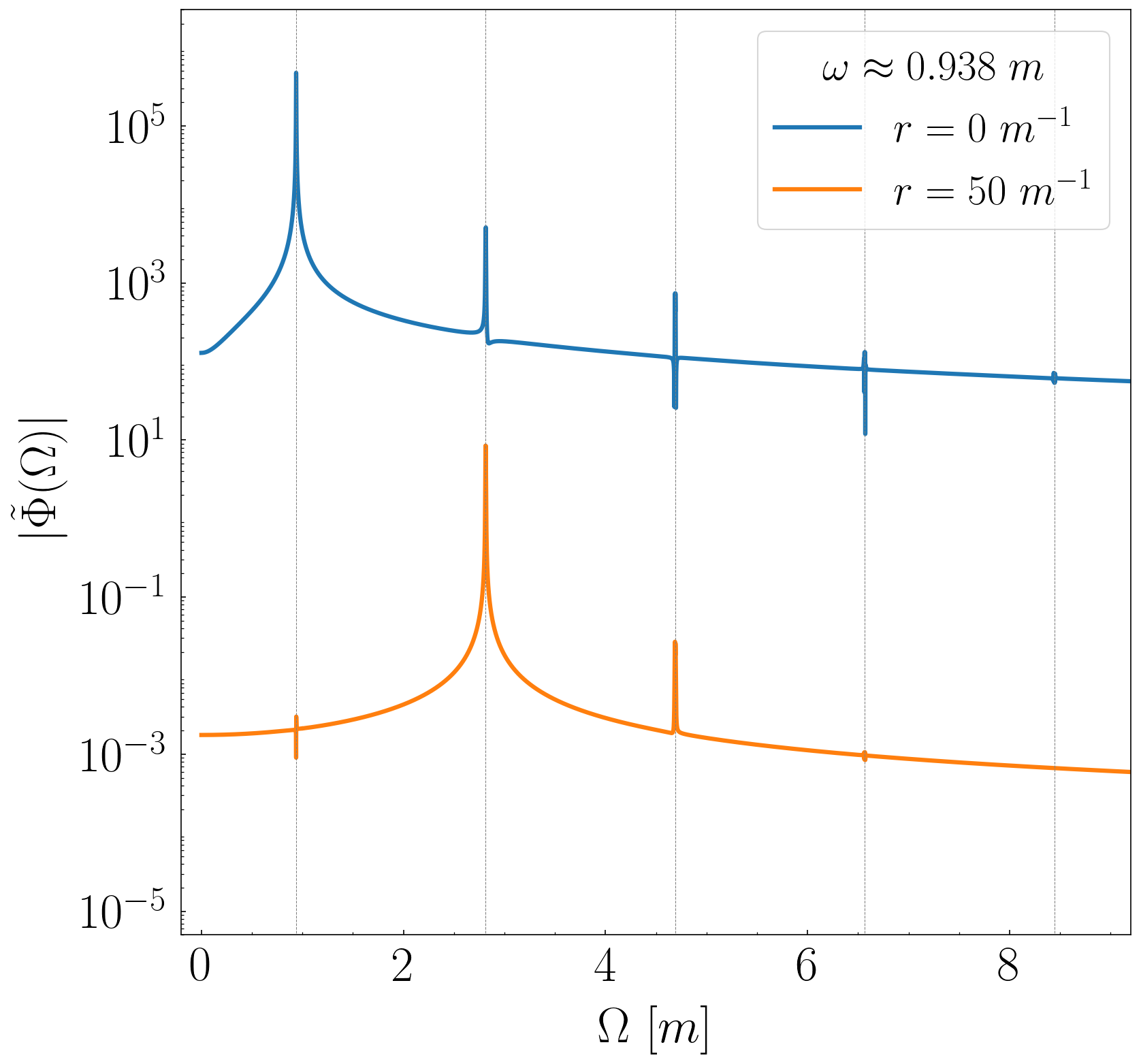}
	\end{minipage}
	\begin{minipage}{0.496\linewidth}
		\includegraphics[width=\linewidth]{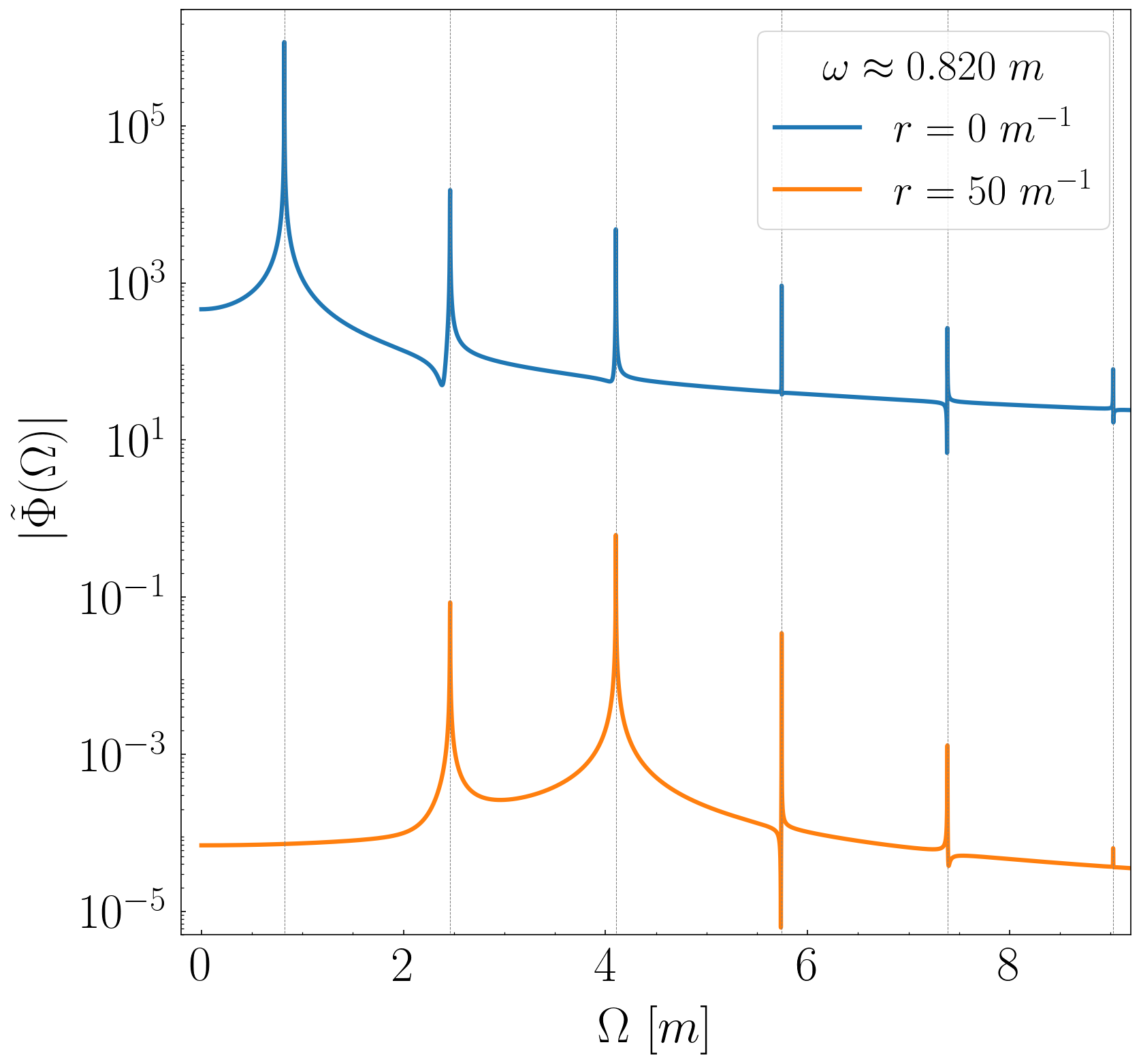}
	\end{minipage}
	\caption{Fourier analysis of the field amplitude at the centre of the oscillon $r=0m^{-1}$ (blue) and far from the center where radiation dominates (orange) [units are arbitrary on the vertical axes]. For both panels, note that the frequency content of the oscillon is dominated by a single fundamental frequency $\omega$, although higher harmonics of $\omega$ are present (blue curves). For the left panel, we have chosen $\omega=0.938m$. In this case the radiation content (orange) is dominated by the $3\omega$ mode as expected, with subdominant content in higher multiples of $\omega$. In contrast, we chose $\omega=\omega_{\star}\approx 0.82m$ for the right panel which is the location of the dip in the decay rate in Fig.~\ref{fig:tanhfreqenergyamp}. As expected, in this case, the $3\omega$ mode is subdominant in the radiation, with the $5\omega$ mode determining the decay rate. These plots provide a verification of our underlying assumptions and confirm the results of our analytic calculation.}
	\label{fig:tanhrootfourieranalysis}
\end{figure}
As our first example, we consider a $\alpha$-attractor T-model from conformal chaotic inflation \cite{Kallosh:2013hoa}, i.e.
\begin{align}
V(\phi)=\frac{m^2M^2}{2} \tanh ^{2}\frac{\phi}{M} ~.
\end{align}
The numerical and analytical results for the field amplitude, energy and decay rate as a function of $\omega$ are presented in Fig.~\ref{fig:tanhfreqenergyamp}.
\\ \\
\noindent{\bf Amplitude and Energy}: In the left panel of Fig.~\ref{fig:tanhfreqenergyamp}, we show the central amplitude and total energy of the oscillon configurations as a function of $\omega$. Note that the amplitudes $\Phi(r=0)/M\gtrsim \mathcal{O}[1]$. The upper-limit of the frequency corresponds to $\omega_{\rm crit}$, above which the oscillons are unstable against long-wavelength perturbations. The black dots indicate the numerically obtained energies and amplitudes as the configurations evolve from low to high $\omega$. The agreement between the colored lines (analytic) and the black dots (numerical) indicates that our single frequency ansatz works reasonably well in the range displayed -- conservatively, it is consistent with the numerical solutions at a few~\% level.
\\ \\
\noindent{\bf Decay Rate}: In the right panel of Fig.~\ref{fig:tanhfreqenergyamp}, we show the numerically calculated decay rate (black dots) as the oscillon evolves with time (from low to high $\omega$) until its eventual demise at $\omega=\omega_{\rm crit}$ at the right edge of the panel. Notice the significant ``dip" in decay rate around $\omega_\star\approx0.82m$. The solid red line shows that most of the lifetime of the oscillons is spent in the dip. We compare these numerically obtained results with the analytic expectation of our calculations. 
\begin{figure}[t]
	\centering
	\begin{minipage}{0.45\linewidth}
		\includegraphics[width=\linewidth]{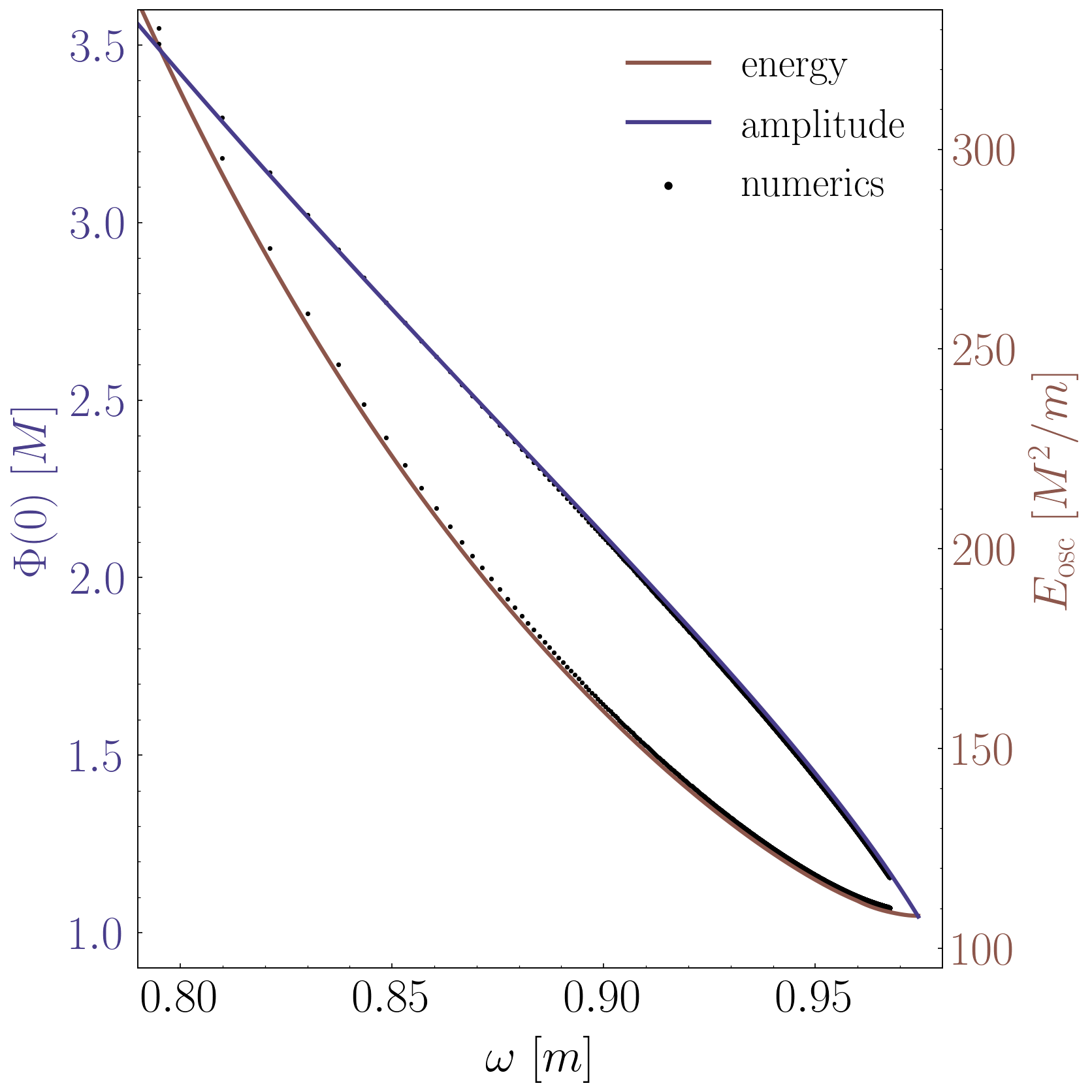}
	\end{minipage}
	\qquad
	\begin{minipage}{0.45\linewidth}
		\includegraphics[width=\linewidth]{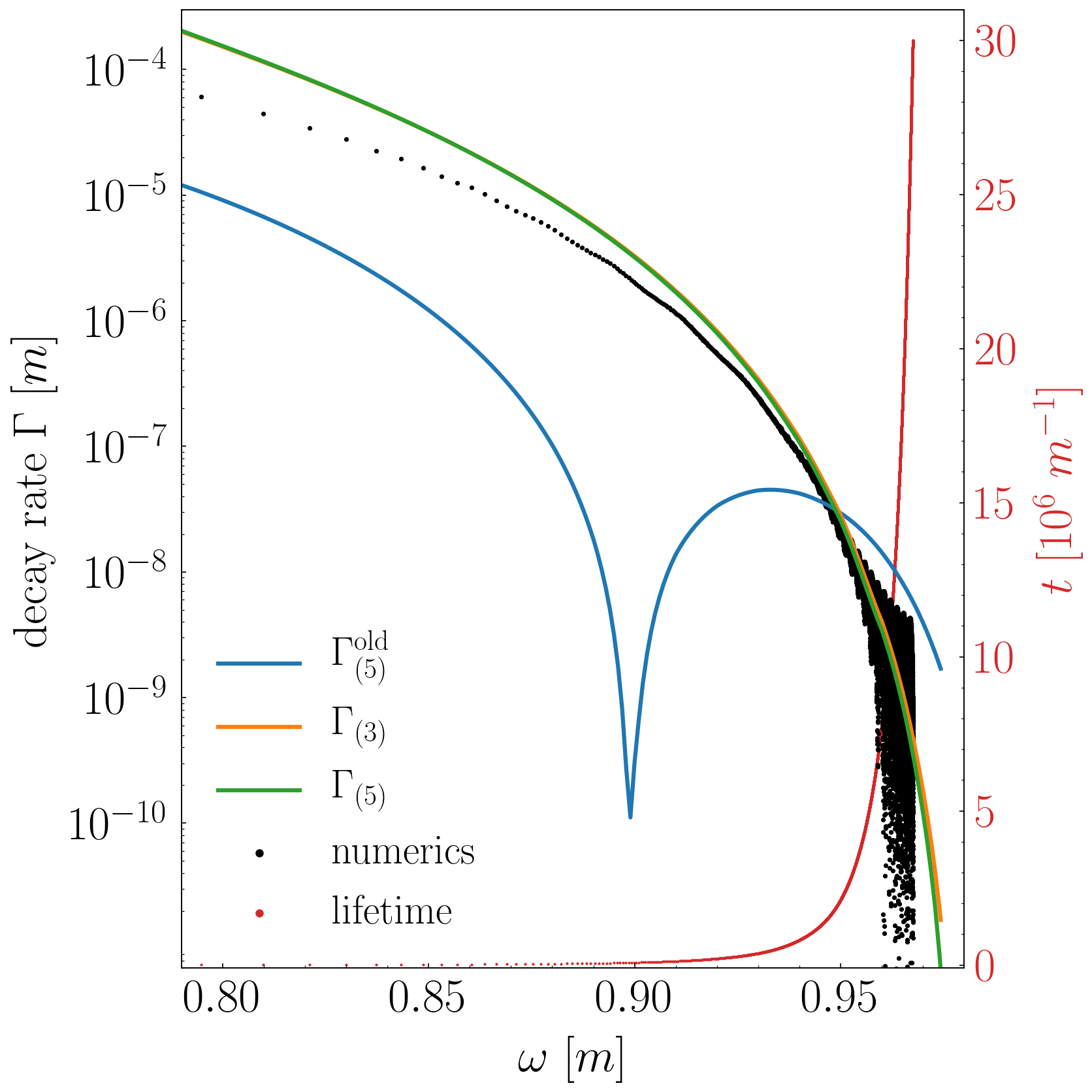}
	\end{minipage}
	\caption{\small $V(\phi)=(1/2)m^2M^2\ln(1+\phi^2/M^2)$: For a general description, see the caption of Fig.~\ref{fig:tanhfreqenergyamp}. Note that the numerics and our analytic calculation match exceptionally well both for the decay rates (right panel) as well as the amplitude and energy (left panel) of oscillons. Note that ignoring the effective mass (blue curve, right panel) incorrectly predicts a dip in the decay rate which is not observed in the numerical calculation (and is in general a bad estimate of the numerics). Also note that the lifetime of this oscillon is larger than our maximum programming time, i.e. $t_{\rm max}=3\times 10^7m^{-1}$. We expect the oscillon to collapse after it reaches $\omega_{\rm crit}\approx 0.974m$.}
	\label{fig:monodromy0freqenergyamp}
\end{figure}

Note that $\Gamma_{(3)}$ (orange curve), where radiation modes with frequency $3\omega$ were included, beautifully captures the location of the dip in $\Gamma$ as a function of $\omega$. In particular, $\tsS_3(\kappa_3)=0$ at $\omega_\star\approx0.82m$ (and hence $\Gamma_{(3)}=\Gamma_3=0$). The $\Gamma_{(5)}=\Gamma_3+\Gamma_5$ calculation (green) barely corrects the $\Gamma_{(3)}$ anywhere, except in the dip, making the decay rate small but finite there. This is to be expected. As we discussed in Sec.~\ref{sec:decay_rate}, we expect $[\tsS_3(\kappa_3)]^2\gg [\tsS_5(\kappa_5)]^2$, except when $[\tsS_3(\kappa_3)]^2$ vanishes. 

For comparison, we also show $\Gamma^{\rm old}_{(5)}$ (blue curve) where the effective mass is ignored. Our calculated $\Gamma_{(5)}$ (green) is a significant improvement (by orders of magnitude) compared to $\Gamma^{\rm old}_{(5)}$, and we correctly capture the dip in the decay rate.  We note, however, that even our calculation overestimates the numerically obtained decay rates. 
\\ \\
\noindent{\bf Frequency Content}: It is useful to calculate frequency content of the oscillon as well as the radiation -- this calculation allows us to verify some of the assumptions inherent in our analytic calculation. We take the Fourier transform of $\phi(t,r=0)$ and $\phi(t,r=r_{\rm rad})$. We provide these Fourier transforms for $\omega=0.938m$ as well as $\omega=\omega_\star\approx 0.82m$ in Fig.~\ref{fig:tanhrootfourieranalysis}. Consistent with our assumptions, note that a single frequency does dominate the profile near the origin (blue). Similarly, for the radiation (orange), there is a clear hierarchy of power in multiples of $\omega$. This hierarchy is broken at the dip for the $\omega_\star\approx 0.82m$ case, with $5\omega$ contribution becoming larger than the $3\omega$ one. We caution that given the finite window for the Fourier transform and numerical uncertainties, the absolute amplitude of the peaks are not quite robust, however, the trends can be trusted.

\subsection{The Logarithmic Potential}
If we take the limit $q\rightarrow 0$ in the general potential \eqref{eq:PotGen}, we have
\Beq\label{eq:log_potential}
V(\phi)=\frac{m^2M^2}{2}\ln \left(1+\frac{\phi^2}{M^2}\right)\,.
\Eeq
In Fig.~\ref{fig:monodromy0freqenergyamp}, we show the comparison between the analytical and numerical results for this potential. Apart from the excellent match between theory and numerics, it is worth noting that the numerics does not show any non-monotonic behavior in the decay rate. Our analytics agree with this behavior (green and orange curves). However, ignoring the effective mass incorrectly predicts the existence of a dip, and provides a poor match for the numerics in general even at large $\omega$. Note that we did not simulate the eventual demise of these oscillons -- their lifetime is longer than $3\times 10^{7}m^{-1}$, and could be longer (but finite).
\subsection{The Axion-Monodromy Potential}
\begin{figure}[t]
	\centering
	\begin{minipage}{0.45\linewidth}
		\includegraphics[width=\linewidth]{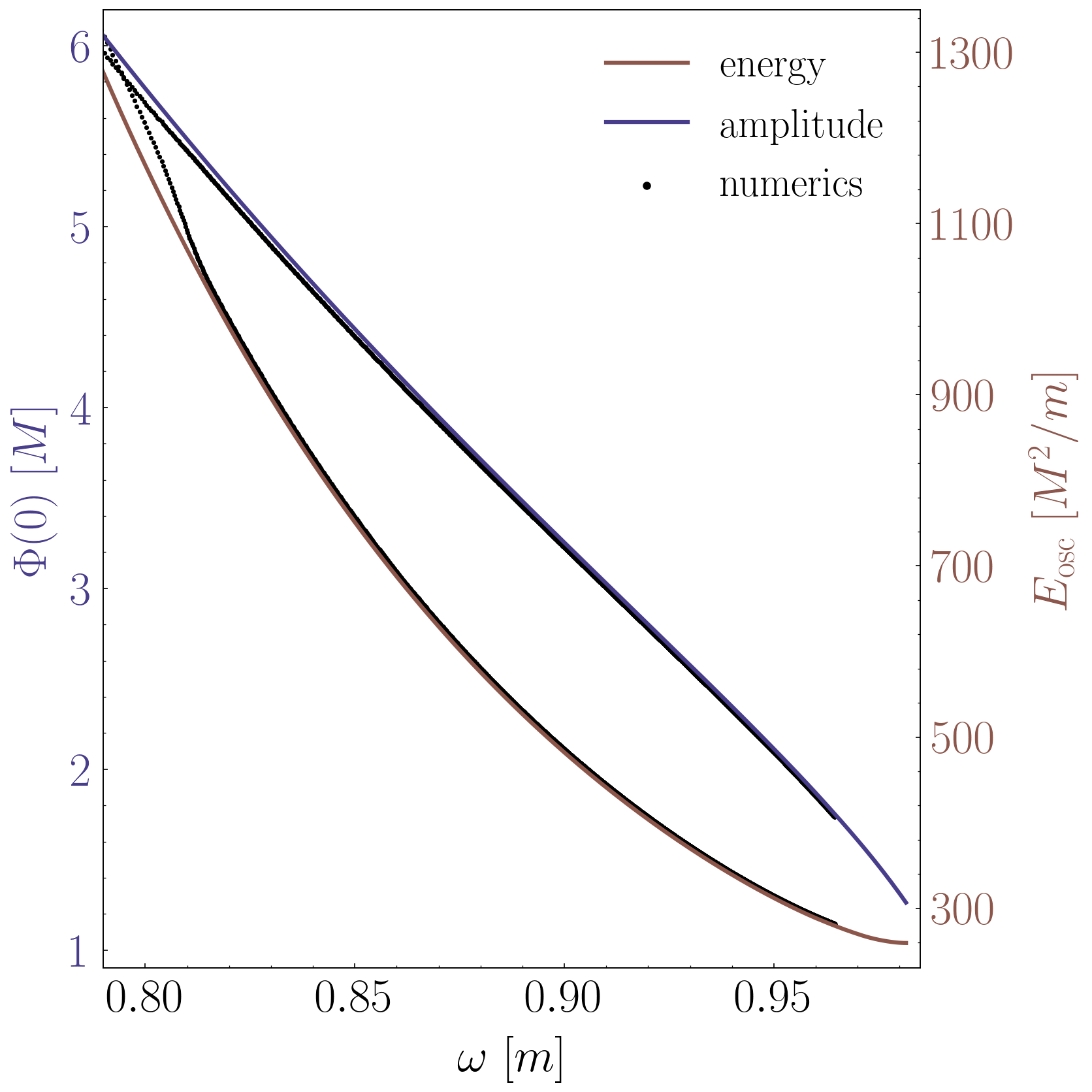}
	\end{minipage}
	\qquad
	\begin{minipage}{0.45\linewidth}
		\includegraphics[width=\linewidth]{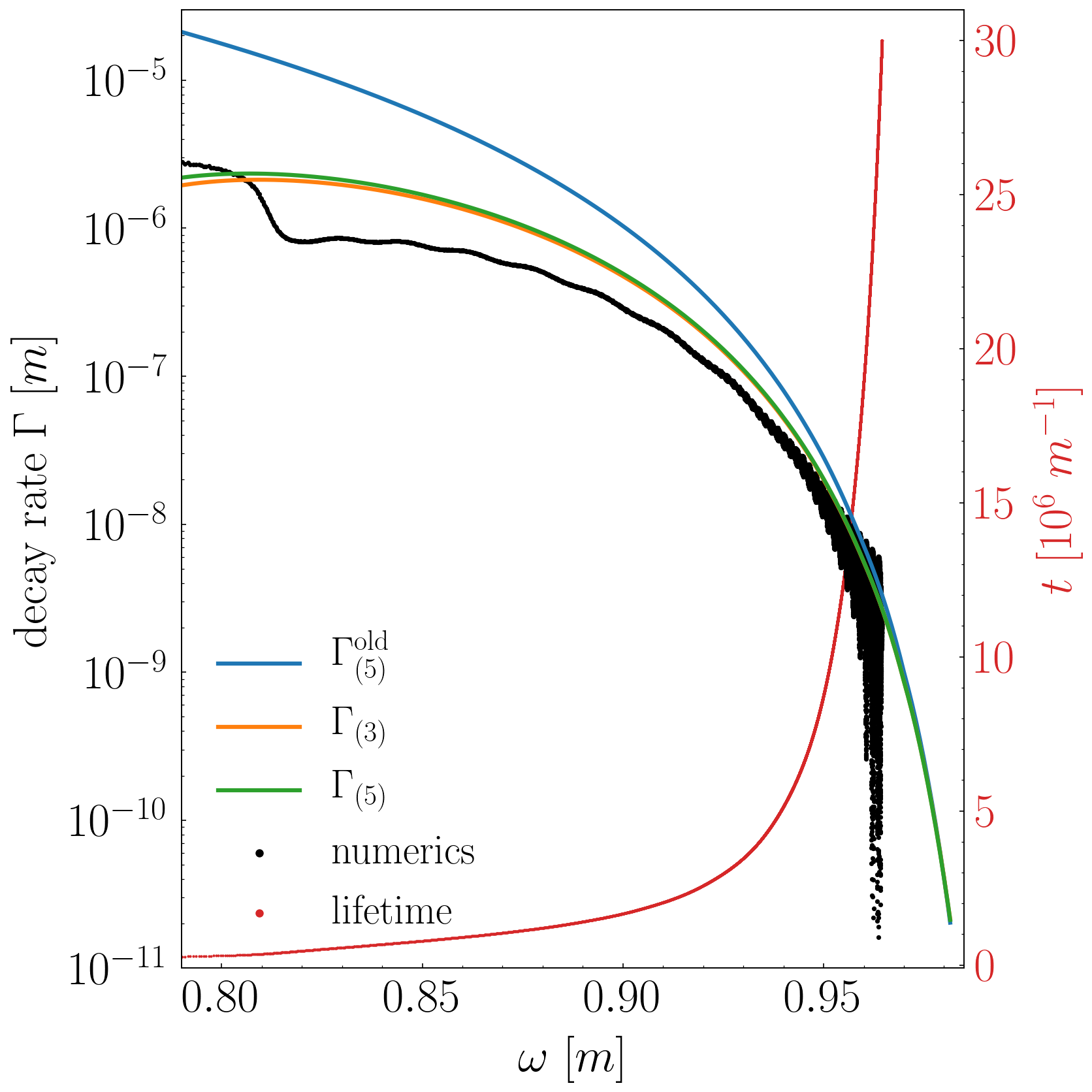}
	\end{minipage}
	\caption{\small $V(\phi)=m^2M^2\left[\sqrt{1+\phi^2/M^2}-1\right]$: For general description, see the caption of Fig.~\ref{fig:tanhfreqenergyamp}. Once again, the analytics and numerics agree quite well. In this case the behavior of the decay rate is monotonic. Our estimates (orange and green curves) are in better agreement with the numerics than the case where the effective mass is ignored (blue). Note that most of the oscillon's lifetime arises from the configuration with a frequency close to the $\omega_{\rm crit}\approx 0.982m$, with a lifetime that is longer than $3\times 10^{7}\,m^{-1}$ (potentially much longer). For all cases where the behavior of the decay rate is monotonic up to the critical frequency, we expect that the oscillon tends to spend most of its lifetime near the critical frequency. }
\label{fig:squarerootfreqenergyamp}
\end{figure}

If we take $q=1$ in the general potential \eqref{eq:PotGen}, we have
\Beq\label{eq:monodromy_potential}
V(\phi)=m^2M^2\left[\sqrt{1+\frac{\phi^2}{M^2}}-1\right]\,.
\Eeq
In Fig.~\ref{fig:squarerootfreqenergyamp}, we show the comparison between the analytical and numerical results for this potential. Apart from the excellent match between theory and numerics, it is worth noting that the numerics do not show any non-monotonic behavior in the decay rate. Our analytics agree with this behavior (green and orange curves). Note that, ignoring the effective mass is a poorer fit to the numerical data. Also note that we did not simulate the eventual demise of these oscillons. Their lifetime is longer than $10^{8}m^{-1}$ \cite{Amin:2011hj,Olle:2019kbo}. 

\subsection{The $q=-1$ Potential}
If we take $q=-1$ in the general potential \eqref{eq:PotGen}, we have
\Beq\label{eq:q-1_potential}
V(\phi)=m^2M^2\left[1-\frac{1}{\sqrt{1+\phi^2/M^2}}\right]\,.
\Eeq
Note that similar to the hyperbolic tangent potential, this potential asymptotes to a constant $m^2M^2$ when $\phi/M\gg 1$. Using $\tsS(\kappa_3)=0$, we can predict that there is a dip in the decay rate at $\omega_{\star}\approx 0.929\,m$. This corresponds to a central oscillon amplitude of $\Phi(r=0)\approx 1.401\,M$. A comparison with the numerical results shown of Ref. \cite{Olle:2019kbo}, shows excellent agreement between our analytic estimates for the amplitude of the oscillon in the dip and its frequency, and their numerical results. The authors in \cite{Olle:2019kbo} showed that the lifetimes are $\gtrsim 6\times 10^8\,m^{-1}$.  We do not repeat their numerical analysis here.

\subsection{The $\phi^6$ potential}
To connect with earlier work, we also investigated
\Beq\label{eq:phi6}
V(\phi)=\frac{1}{2}m^2\phi^2-\frac{\lambda}{4} \phi^4+\frac{g}{6} \phi^6\,,
\Eeq
where $\lambda,g>0$. This potential supports stable oscillons within a finite range of amplitudes (and frequencies), including flat-top oscillons \cite{Amin:2010jq}. Unlike the other cases considered in this paper,  this potential is steeper than quadratic asymptotically, and leads to some simplifications in the calculations since the potential is a low-order polynomial. 

Analytical and numerical results confirm that the decay rate is non-monotonic, and shows multiple dips indicating the existing of multiple, long-lived oscillon configurations. In this case, our analytical calculations do not show an improvement compared to the case where the effective mass is ignored (it shifts the dips around the numerically obtained values).  Since this case (ignoring the effective mass) has been explored in detail in \cite{Mukaida:2016hwd,Ibe:2019vyo}, we do not repeat the analysis here, however it was the analysis in those publications that has motivated our present work. 


\subsection{The cosine potential}
Oscillons in the cosine potential
\Beq\label{eq:phi6}
V(\phi)=m^2M^2\left[1-\cos\left(\frac{\phi}{M}\right)\right],
\Eeq
might be of particular interest, since such a potential appears in many systems with a discrete shift symmetry, including QCD axions. Stable oscillon solutions exist for this potential in 1+1 spatial dimensions, however, the situation is different in $3+1$ dimensions. With our techniques, we find that the lifetime of oscillons in the cosine potential is quite short: $\lesssim \mathcal{O}[10^3]\,m^{-1}$. Such relatively short lifetimes are consistent with earlier studies (see, for example, \cite{Piette_1998}). Note that longer lifetimes can be found when including gravitational effects in certain limits, which we discuss further in the next session.


\section{Unresolved Issues and Future Directions}\label{sec:future}
\noindent{\bf Decay Rate Over-Estimate}: For all the cases we have considered, we have found that our analytic predictions for the decay rate typically over-estimates the numerically calculated decay rate at large amplitudes. Including radiation at additional multiples of $\omega$ in the decay rate calculation does not help, since they are all positive definite (and small) contributions. A way to improve our calculation would be to include multiple frequencies in the profile itself; this could potentially bring the analytical results in even better agreement with the numerics. 

We note that the over-estimate in the `dips' in $\Gamma$ prevents us from obtaining accurate lifetime estimates in cases where such dips exist.
\\ \\
\noindent{\bf Discrepancies at  Large Amplitudes}: In most of the cases we have considered, our decay rates deviate more and more from the numerical ones at sufficiently large amplitude ($\Phi(0)\gtrsim {\rm few} \times M$). This is likely due to the presence of multiple frequencies (not necessarily multiples of the fundamental frequency) in the oscillon, as well as a breakdown of the hierarchy between lower and higher frequency modes of the radiation. We are unable to provide a systematic criterion for the amplitude or frequency at which we should stop trusting our approach (or a bound on the errors) beyond the rough guide that our approach works well when $\Phi(0)\lesssim {\rm few}\times M$. The energy of the oscillon calculated using the single frequency assumption continues to provide a reasonably good approximation to the numerically evaluated energy even at large central amplitudes, however, the decay rate becomes less reliable.\footnote{We also note that at large amplitudes, a very low momentum mode with frequency $\Omega =(1+\epsilon)m$ with $0<\epsilon\ll 1$ appears in a Fourier analysis. These might be related to low momentum, free particles which are neither part of the oscillon, nor of the outgoing radiation. These are typically seen to be created when an initial field configuration is settling down to an oscillon configuration.}

These concerns are less critical than one might imagine since oscillons that form from cosmological initial conditions typically have amplitude $\sim {\rm few}\times M$. Properties of such oscillons are correctly captured by our calculational framework.
\\ \\
\noindent{\bf Asymmetric Potentials}: We have focused on $V(\phi)=V(-\phi)$ potentials in this paper. However, lifetimes of oscillons in potentials with $V(\phi)\ne V(-\phi)$, for example, $V(\phi)=(1/2)\phi^2+(\lambda_3/3)\phi^3+(\lambda_4/4) \phi^4$ have been of interest from the early papers \cite{Gleiser:1993pt,Copeland:1995fq} (and recently, see for example \cite{Antusch:2019qrr,Gleiser:2019rvw}). We found that our techniques can be applied in principle to these potentials, however the application is complicated because to capture the background profile, we require more than one mode \cite{Saffin:2006yk}. This complication feeds in to the radiation calculation. These types of potential appear naturally in spontaneous symmetry breaking contexts, and are worth exploring further. 

Infinite lifetimes for such potentials were conjectured in \cite{Honda:2001xg,Gleiser:2019rvw} based on intriguing spikes in lifetimes as a function of the initial radius of a Gaussian profile (with a fixed initial amplitude).\footnote{Numerically, the authors in \cite{Gleiser:2019rvw} found that the lifetime for these ``resonant" configurations was found to be larger by $\sim 10\%$ compared to the non-resonant ones (with the maximum total lifetime observed being $\sim 10^4$ oscillations).} As discussed in Sec.~\ref{sec:decay_rate}, our analysis of classical decay rates in symmetric potentials does not lend support to the idea of infinite lifetimes. We found that even when radiation of a given frequency vanishes for some special oscillon configuration, it does not do so for higher multiples of that frequency. Nevertheless, it would be useful, though non-trivial, to investigate asymmetric double-well potentials using our techniques.
\\ \\
\noindent{\bf Quantum Effects/ Short-Wavelength Stability Analysis}: We have carried out a calculation of the classical decay rate. As discussed in \cite{Hertzberg:2010yz,Saffin:2014yka} (also see \cite{Olle:2019skb}), the presence of quantum fluctuations might reduce the lifetime. Heuristically, the exponential suppression in the decay rates arising from $[\tsS_j(\kappa_j)]^2$ might be absent when one includes quantum fluctuations -- although the analysis in our large/moderate amplitude limit remains to be done carefully. 

If we include fluctuations (mimicking zero-point fluctuations) only in the initial state, then the above analysis is intimately related to analysing the stability of oscillons to short wavelength fluctuations. This calculation is feasible using a Floquet-type analysis with coupled Fourier modes \cite{Hertzberg:2010yz}. We carried out such a linear instability analysis for the spatially inhomogeneous oscillon background in a limited number of cases and found that short-wavelength instabilities do exist analytically (but without significant features at the dips in decay rates). However, when exploring their impact numerically, their fate in the nonlinear regime is not completely clear. We leave the detailed study of such instabilities for future work. 
\\ \\
\noindent{\bf Gravitational Effects}: We have ignored gravity in our calculations. Aspects of gravitational effects  on the decay rates were considered in (for example) \cite{Fodor:2009kg,Eby:2015hyx, Eby:2017azn, Eby:2017teq, Visinelli:2017ooc,Eby:2018ufi,Eby:2019ntd} in the weak field limit, with a number of authors focussing on the cosine potential relevant for axions. For the cases we have considered here, it is quite plausible that inclusion of gravitational effects will allow for even longer-lived configurations. However, it is also possible that in some cases gravity will destabilize the oscillons, making them collapse or disperse quickly. For recent numerical relativity studies of the fate of oscillons in the strong field gravity regime, see \cite{Helfer:2016ljl,Muia:2019coe}. We leave a detailed analytic calculation of gravitational effects on the decay rates for future work. 

\section{Summary}\label{sec:conclusion}

Oscillons have been known to have exceptionally long lifetimes in spite of the fact that there is no strictly conserved charge (unlike Q-balls). In this paper, we have provided an improved framework for calculating the small decay rates of oscillons. The key reason for the improvement is the inclusion of a spacetime dependent effective mass in the equations of motion of the radiation modes.

Our calculations are based on two important assumptions: (1) The oscillon is dominated by a single temporal frequency $\omega$. (2) The radiation modes are excited at multiples of $\omega$, with a lower frequency radiation typically dominating over higher frequency radiation (apart from some exceptionally long-lived configurations discussed further below). For practical reasons, we only included finite number of radiation modes. We checked these assumptions a-posteriori from our numerical simulations.

The main improvements in the framework and corresponding results are as follows:
\begin{enumerate}
\item By systematically including the spacetime-dependent effective mass in the equations describing outgoing scalar radiation, we demonstrated that its inclusion is essential for understanding the presence or absence of critical features in the decay rate as a function of oscillon parameters. We showed that these critical features arise because for certain oscillon configurations, the $3\omega$ radiation mode that typically determines the decay rate is unusually suppressed. This leads to exceptionally long-lived  configurations. Their lifetimes are still finite since radiation at higher frequencies does not typically vanish for these configurations. In the absence of such features, we find that the decay rate  is monotonically decreasing with the fundamental frequency of the oscillon.  The decay rate can still become extremely small before a rapid dissipation of the oscillon near $\omega_{\rm crit}$. 

\item We compared our numerical simulations and analytic results for the decay rates. For the numerical simulations, we started with a large amplitude (low frequency) oscillon and allowed it to evolve by emitting scalar radiation. The energy of the configuration decreases, and the field passes through different oscillon configurations with the frequency always increasing with time. The analytic comparison is done with these numerical results, treating the field configurations as quasi-stationary.

\item The improved decay rate estimates provided in this paper are in better agreement with the full numerical simulations, resulting an improvement (compared to existing techniques) by many orders of magnitude (for example, see Fig.~\ref{fig:tanhfreqenergyamp} and Fig.~\ref{fig:monodromy0freqenergyamp}).

\item We emphasize that we are able to capture the salient features of the oscillon profiles and decay rates for large amplitude oscillons, and in flattened potentials, which are particularly relevant in cosmological applications. Most of the earlier literature is focused on small amplitude oscillons, or on models with potentials that are well approximated by a low-order polynomial.\footnote{We also studied polynomial potentials ($\phi^6$ potentials) for comparison with earlier work \cite{Mukaida:2016hwd,Ibe:2019vyo}, but did not find any significant improvement over those results when we included the effective mass term in the analysis.}  Our ability to predict the exceptionally long-lived states (classical lifetimes $\gtrsim 10^8 m^{-1}$)  would be useful in knowing which oscillons would survive the longest in the case where an ensemble of such objects would form from cosmological initial conditions. 

\end{enumerate}

We discussed some of the shortcomings of our approach including typical over-estimates of the decay rates. In addition, we ignored the impact of quantum fluctuations on the decay rates (which will likely reduce lifetimes) as well as the impact of gravity (which might potentially increase lifetimes). We also pointed to some difficulties of our approach when dealing with asymmetric potentials. Given the present success of our approach, we hope that it will be worthwhile for others (and us) to build on our formalism and take on these challenges.

\section{Acknowledgements}
MA, EJC and KL acknowledge support from the Durham IPPP Visiting Academics (DIVA) program. MA and HYZ are supported by a NASA ATP theory grant NASA-ATP Grant No. 80NSSC20K0518. MA, EJC, HYZ and KL acknowledge the hospitality of the KITP during the running of the wonderful workshop, ``From Inflation to the Hot Big Bang" supported in part by the National Science Foundation under Grant No. NSF PHY-1748958, where part of this work was carried out.  EJC and PMS acknowledge support from STFC grant ST/P000703/1. We would also like to thank M. Cicoli, F. Muia, L. Street and R. Wijewardhana for useful feedback.

\bibliographystyle{utphys}
\phantomsection 
\addcontentsline{toc}{section}{References} 
\bibliography{Reference}

\appendix
\section{Solution of Perturbations}\label{sec:derivation_xi}

Ignoring the $\mathcal{E}_j$ terms in \eqref{eq:xi_master}, the perturbation equations become
\begin{align}
(\kappa_j^2 + \nabla^2) \xi_j(\b x) = -S_j(\b x) ~,
\end{align}
which is the Helmholtz equation. It is easier to solve it in Fourier space, i.e.
\begin{align}
\til \xi_j(\b k) = \frac{\til S_j(\b k)}{k^2 - \kappa_j^2} ~,
\end{align}
then assuming the spherical symmetry, the solution in coordinate space becomes
\Beq
\xi_j(r) &= -\frac{i}{8\pi^2 r}\int_{-\infty}^\infty dk \frac{k\til S_j(k)}{k^2-\kappa_j^2} \( e^{ikr}-e^{-ikr} \) \\
&= - \frac{1}{4\pi r} \int_0^\infty dr' ~ r' S_j(r') \int_{-\infty}^{\infty} \frac{dk}{k^2-\kappa_j^2} \[ e^{ik(r+r')}-e^{ik(r-r')}  +  e^{-ik(r+r')} - e^{-ik(r-r')} \] \\
\label{xi_solution}
&= \frac{\cos(\kappa_j r)}{\kappa_j r} \int_0^r dr' ~ S_j(r')~ r' \sin(\kappa_j r') + \frac{\sin(\kappa_j r)}{\kappa_j r} \int_r^\infty dr' ~ S_j(r')~ r' \cos(\kappa_j r') ~,
\Eeq
where in the first equal sign we have doubled the domain of integration based on the fact that the integral is even in $k$. In the second equal sign we have plugged the Fourier transform of $\til S_j(k)$. In the third line we have evaluated the integrals using:
\begin{align}\label{residue_real_axis}
\int_{-\infty}^{\infty} f(x) \mathrm{d} x 
= \pm2 \pi \mathrm{i} \sum_{j=1}^{n} \operatorname{Res} f(z_j)\pm \pi \mathrm{i} \sum_{j=1}^{m}\operatorname{Res} f(\alpha_j) ~,
\end{align}
where $z_j$ are singularities on upper(+)/lower(-) half of the complex plane and $\alpha_j$ are singularities on the real axis. Note that this result is called \emph{Cauchy principal value} and is independent of the choice contour.

 If $r\rightarrow\infty$, our solution reaches a simple form
\begin{align}
\xi_j(r) = \frac{1}{4\pi r} \til S_j(\kappa_j) \cos(\kappa_j r) ~.
\end{align}
Including the time part and ignoring the contributions of ingoing waves, we obtain
\begin{align}
\xi(t,r) = \frac{1}{8 \pi r} \sum_{j} \tilde{S}_{j}\left(\kappa_{j}\right) \cos \left(\kappa_{j} r-\omega_{j} t\right) ~.
\end{align}
This differs from the expression used in eq.~\eqref{eq:xi_rad} by a factor of $2$. 

Instead of the above approach, we could have used a retarded Green's function of the Klein-Gordon operator  to solve \eqref{EOM_perturbation} (with an associated $i \epsilon$ prescription). This approach would have avoided the seemingly unreasonable expansion of perturbations in standing cosine series \eqref{perturbation_expansion}.  The final results of these two methods simply differ by a factor of $2$. The use of the Cauchy principal value in our standing wave approach leads to half the value since no $i\epsilon$ prescription was chosen \cite{arfken1999mathematical,landau2008quantum}.

\end{document}